\documentclass[10pt,conference]{IEEEtran}
\IEEEoverridecommandlockouts
\usepackage{bm}
\usepackage{graphics}
\usepackage{enumerate}
\usepackage{siunitx}
\usepackage{amsmath}
\usepackage{amsfonts}
\usepackage{amsfonts}       % blackboard math symbols

\usepackage{amssymb}
\usepackage{xcolor,soul}
\sethlcolor{lightgray}
\usepackage{multirow}
\usepackage{graphicx}
\usepackage[hyphens]{url}
\usepackage{color}

\begin{document}

%\title{Uncertainty in Performance Prediction for Configurable Software Systems via Bayesian Deep Learning}
\title{Uncertainty-Aware Performance Prediction for Highly Configurable Software Systems via Bayesian Neural Networks}

\author{\IEEEauthorblockN{Huong Ha}
\IEEEauthorblockA{\textit{RMIT University} \\
Melbourne, Australia \\
huong.ha@rmit.edu.au}
\and
\IEEEauthorblockN{Zongwen Fan}
\IEEEauthorblockA{\textit{Huaqiao University} \\
Xiamen, China \\
zongwen.fan@hqu.edu.cn}
\and
\IEEEauthorblockN{Hongyu Zhang}
\IEEEauthorblockA{\textit{The University of Newcastle} \\
Newcastle, Australia \\
hongyu.zhang@newcastle.edu.au}
}

% \author{Huong Ha}
% \affiliation{
%   \institution{RMIT University}
%   \city{Melbourne}
%   \state{VIC}
%   \country{Australia}
%   \postcode{3000}
% }
% \email{huong.ha@rmit.edu.au}
%
% \author{Zongwen Fan}
% \affiliation{
%   \institution{Huaqiao University}
%   \city{Xiamen}
%   \state{Fujian}
%   \country{China}
%   \postcode{361021}
% }
% \email{zongwen.fan@hqu.edu.cn}
%
% \author{Hongyu Zhang}
% % \authornote{Both authors contributed equally to this research.}
% \affiliation{
%   \institution{The University of Newcastle}
%   \city{Callaghan}
%   \state{NSW}
%   \country{Australia}
%   \postcode{2308}
% }
% \email{hongyu.zhang@newcastle.edu.au}

\pagestyle{plain}
\maketitle

\begin{abstract}
Configurable software systems are employed in many important application domains. Understanding the performance of the systems under all configurations is critical to prevent potential performance issues caused by misconfiguration. However, as the number of configurations can be prohibitively large, it is not possible to measure the system performance under all configurations. Thus, a common approach is to build a prediction model from a limited measurement data to predict the performance of all configurations as scalar values. However, it has been pointed out that there are different sources of uncertainty coming from the data collection or the modeling process, which can make the scalar predictions not certainly accurate. To address this problem, we propose a Bayesian deep learning based method, namely \textbf{BDLPerf}, that can incorporate uncertainty into the prediction model. \textbf{BDLPerf} can provide both scalar predictions for configurations' performance and the corresponding confidence intervals of these scalar predictions. We also develop a novel uncertainty calibration technique to ensure the reliability of the confidence intervals generated by a Bayesian prediction model. Finally, we suggest an efficient hyperparameter tuning technique so as to train the prediction model within a reasonable amount of time whilst achieving high accuracy. Our experimental results on 10 real-world systems show that \textbf{BDLPerf} achieves higher accuracy than existing approaches, in both scalar performance prediction and confidence interval estimation.
 
%  To develop \textbf{BDLPerf}, we first construct a sparse Bayesian neural network (BNN) based on the variational inference method.

\end{abstract}

% <concept_id>10011007.10011006.10011071</concept_id>
% <concept_desc>Software and its engineering~Software configuration management and version control systems</concept_desc>

% \ccsdesc[500]{Software and its engineering~Software configuration management}

% \keywords{Configurable Software Systems, Performance Prediction, Bayesian Deep Learning, Uncertainty Estimate}


\section{Introduction} \label{sec:introduction}

Many complex modern software systems, such as database management systems, image processing tools, video encoders and code optimizers are highly configurable. They provide different configuration options for users to customize to meet their specific requirements, and thus improving the usability of the systems. Different configurations could lead to different performance, therefore it is necessary to understand the performance of a system under all configurations before configuring and deploying the system. This helps users to make rational decisions in configurations and avoid misconfiguration, which could lead to system performance issues, negatively affecting their experience. In fact, some empirical studies found that 59 percent of performance issues are related to configuration errors \cite{Han2016PerfEmpirical, Velez2022PerfDebug}, and it is well known that performance issues can damage user experience \cite{Han2016PerfEmpirical, He2020PerfBugs, Li2016EngeryOptimize, Song2017PerfDiagnosis, Wilke2013EngeryMobileApps, Velez2022PerfDebug}. %, or worse, can even cause lawsuits to businesses \cite{HollisterPerflawsuit2020}.

In reality, it is almost impossible to measure the performance of a configurable software system under all configurations as the number of possible configurations could be prohibitively large even for small-scale systems. For example, the popular file archive utility 7z has 44 configuration options, giving rise to $2^{44}$ possible configurations with approximately 70,000 valid configurations \cite{dorn2020mastering}. To address the problem of the explosion of the number of configurations, researchers aims to build a prediction model from the performance data of a small number of configurations, and then predict the performance of the system under a new configuration \cite{siegmund2015performance, Zhang2015PerfFL, Guo2018Decart, ha2019deepperf, ha2019performance}.

\begin{figure}[h]
% \vspace{-0.2cm}
\centerline{\includegraphics[width=73mm,height=40mm]{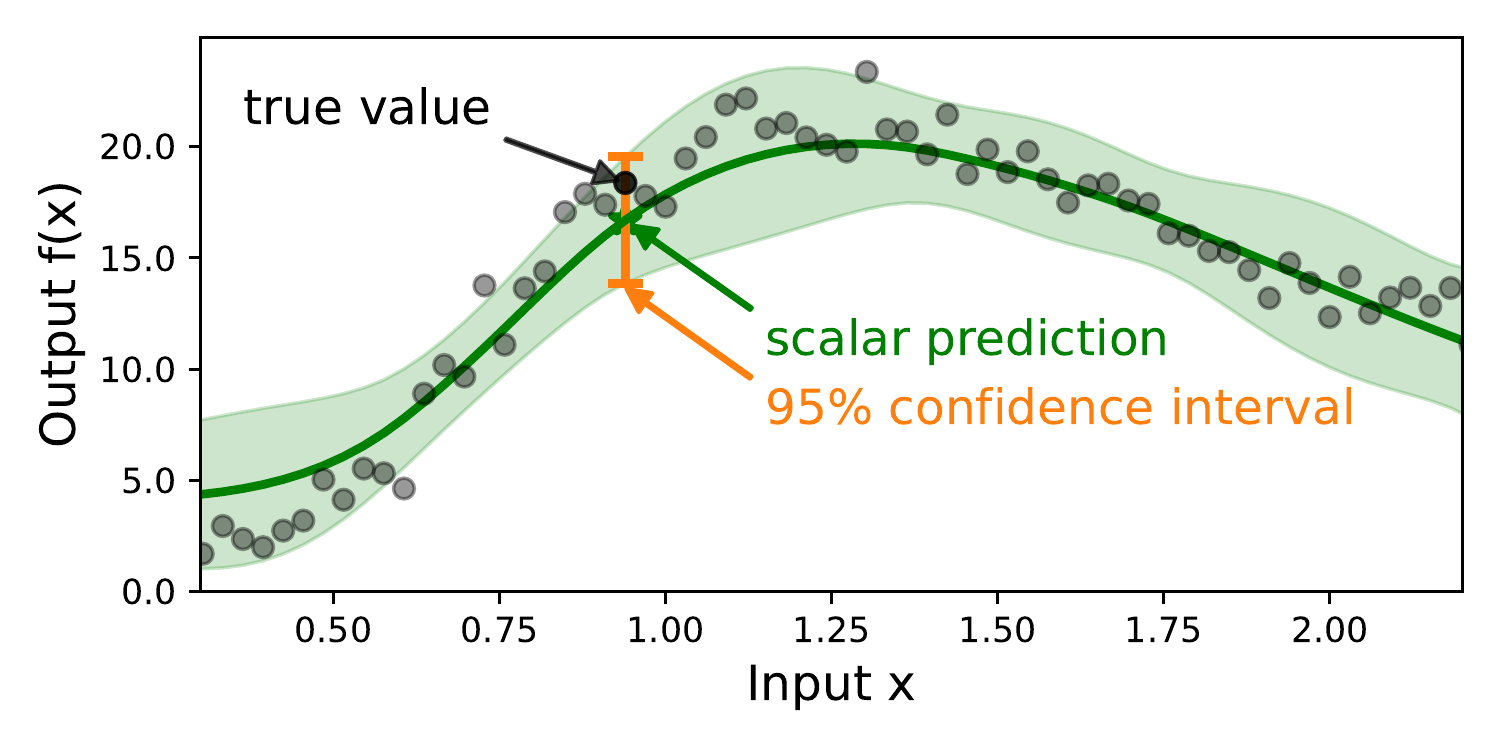}}
\vspace{-0.2cm}
\caption{An illustration of scalar predictions and their confidence intervals on a synthetic data set. Due to different uncertainty factors, the scalar predictions (green line) of a prediction model cannot be 100\% accurate. Confidence intervals (green shaded region) are needed to provide clear expectation boundaries for the scalar predictions.}
%\vspace{-0.2cm}
\label{fig:uncertainty_illustration}
\end{figure}

Recently, it has been suggested that this approach neglects numerous sources of uncertainty, such as the measurement bias, the model representation and learning process, and the limited training data \cite{dorn2020mastering}. Due to these uncertainties, the scalar predictions provided by existing prediction models are highly impossible to be 100\% accurate. A sensible approach is to develop prediction models that can account for these sources of uncertainty, and can provide both scalar predictions and confidence intervals of these scalar predictions \cite{dorn2020mastering}. Figure \ref{fig:uncertainty_illustration} illustrates a simple example of this research problem, where the prediction model provides both scalar predictions (green line) and the corresponding confidence intervals (green shaded region). In practice, confidence intervals help users to manage risk better as it provides the range where the true performance value likely falls in, thus helps users to understand the worst case and best case scenarios when choosing a specific configuration. This is particularly important in safety-critical application domains where many systems are configurable, e.g., automotive systems, cloud-computing systems, cyber-physical systems \cite{KennerPerfSafety2021}.

% (1) Safety, Security, and Configurable Software Systems: A Systematic Mapping Study in SPLC .In practice, as a confidence interval provides the range that the true performance value likely falls in, it helps users to manage the risk when choosing the configuration, avoid potential issues caused by misconfiguration.
% \hy{can describe the importance of the problem and the Figure 1 a bit more, and why an interval is better}
% \hh{emphasize safety issue here if we have the confidence interval}

%Therefore, as proposed in \cite{dorn2020mastering}, 
% \hh{Need to emphasize that all the methods that can provide confidence intervals are Bayesian methods, so can't say because our method is also Bayesian, it's just a simple incremenetal approach compared to Dorn et al.}
To incorporate different sources of uncertainty into a prediction model and generate the confidence intervals, the standard approach is to perform Bayesian inference on the prediction model \cite{Gal2016Uncertainty, Gal2016MCdropout, AbdarUncertainty2021, Lakshminarayanan2017, HullermeierUncertainty2021}. For the configurable software performance prediction problem, recently, Dorn et al.~\cite{dorn2020mastering} proposed to use a linear regression model, and perform Bayesian inference on this model to generate the confidence intervals. However, it is known that linear regression does not work well in the scenarios when the relationship between the system performance and the configuration options are non-linear and complex \cite{Goodfellow2016DL, ha2019deepperf}. In practice, these scenarios generally occur when the configurable software systems are large and complex. Therefore, their technique generally does not work well for complex configurable software systems. 
% \hy{revised a little,pls chk}
% \hy{not just 'overcome', can say we aim at better confidence interval for highly configurable systems, etc}

In this paper, we aim to develop a new method that can generate better confidence intervals for configurable software system performance, especially the complex systems. We suggest to use a deep neural network as the prediction model and develop techniques to perform Bayesian inference on the model in an effective manner given a limited training data. The motivation behind this idea is that deep neural networks are known to be effective in approximating non-linear complex functions \cite{Goodfellow2016DL, Barron1993, Hornik1989, Hornik1991, Funahashi1989}, including highly configurable software systems \cite{ha2019deepperf}. To perform Bayesian inference on the deep neural network, we combine the variational inference \cite{Jordan1999VI} and the ensemble method \cite{Lakshminarayanan2017}, and thus, creating an ensemble of Bayesian neural networks (BNNs) \cite{Mackay1992,Neal1995,Gal2016Uncertainty}. Besides, for a configurable software system, it is known that only a small number of configuration options and their interactions have substantial impact on the system performance~\cite{Jamshidi2017transferlearningperf,siegmund2015performance, Siegmund2012performanceauto}, we then construct sparse BNNs so as to incorporate this knowledge into the model. 

% \hh{maybe highlight the importance of calibration in uncertainty - Dorn et al did not talk about or solve this problem. The issue of calibration can affect any uncertainty generation method not just BNN, i.e. the proposed method can work with the method in Dorn et al.}

In general, Bayesian inference techniques can generate confidence intervals of the scalar predictions, however, it is well known that these confidence intervals are often inaccurate~\cite{Lakshminarayanan2017, Kuleshov2018}. For example, a generated 95\% confidence interval might not contain the ground truth 95\% of time. This issue is particularly severe in the configurable software performance prediction problem where the amount of training data is limited, causing the generated confidence intervals even more inaccurate. To tackle this issue, we therefore propose an uncertainty calibration technique inspired by the Platt scaling technique \cite{Platt99scaling, Kuleshov2018}, but develop it to work with the limited training data scenario. Our uncertainty calibration method is model-diagnostic, i.e. it is general and can be applied to different types of prediction model.

Finally, as with any deep learning methods, hyperparameter tuning is a critical process to ensure the deep networks achieve high prediction accuracy within a reasonable training time. In this work, we also propose an efficient and effective hyperparameter tuning technique via Bayesian Optimization, a powerful black-box global optimization method \cite{Snoek2012}.

To evaluate the performance of our proposed method, \textbf{BDLPerf}, we conduct an extensive set of experiments with ten real-world configurable software systems from different application domains. Our experimental results show that, for both scalar performance prediction and confidence interval estimation, \textbf{BDLPerf} outperforms state-of-the-art techniques on most of the subject systems.

In summary, our key contributions are:
\begin{enumerate}
    \item We propose a Bayesian deep learning model that can generate both scalar predictions and confidence intervals for configurable software performance, and work well with the limited training data scenario.
    \item We develop an effective uncertainty calibration method to calibrate the confidence intervals generated by the Bayesian inference techniques for configurable software performance, and an efficient hyperparameter tuning strategy for our proposed model.
    \item We implement our proposed method, namely \textbf{BDLPerf}, and conduct extensive experiments to evaluate its effectiveness on ten real-world systems and various sampling strategies. Our experimental results show that \textbf{BDLPerf} outperforms existing methods on most subject systems.
\end{enumerate}

% \vspace{-0.5cm}
\section{Background}
\subsection{The Performance Prediction Problem for Configurable Software System}

In general, the performance value of a configurable software system with $n$ configuration options can be expressed as a function  $f(\textbf{x})=f(x_1,x_2,...,x_n):\mathbb{X}\to\mathbb{R}$, where $\mathbb{X}$ is the Cartesian product of the domains of all the configuration options, $x_i \ (i=1,2,\cdots,n)$ is the variable that stores the value of the $i^{th}$ configuration option. The value of variable $x_i$ can be either a Boolean or a real value. Table \ref{table:truthtable_confsystem} shows an example of a configurable software system with $11$ configuration options $x_1, x_2,\dots,x_{10},x_{11}$ and the corresponding performance values of all the configurations.

Given a limited measurement data $\mathcal{D}_{tr}=\{(\textbf{x}_i, y_i)\}_{i=1}^N$, where $y_i = f(\textbf{x}_i) + \epsilon_i$ with $\epsilon_i$ being the corrupted noise (e.g., measurement error), the goal is to train a prediction model that can accurately predict the performance of the configurable software system under new configurations.

\begin{table}[htb]
\begin{center}
% \vspace{-4pt}
\caption{Example of a configurable software system and its performance values.}
%\vspace{-2pt}
\begin{tabular}{ |c|c|c|c|c|c|c|c|c| } 
 \hline
 $x_1$ & $x_2$ & $x_3$ & $\dots$ & $x_8$ & $x_9$ & $x_{10}$ & $x_{11}$ & $f(\textbf{x})$ \\ \hline 
 1 & 1 & 0 & $\dots$ & 10 & 35 & 4 & 2 & 43.62  \\ 
 0 & 1 & 1 & $\dots$ & 12 & 34 & 5 & 4 & 48.13 \\ 
 . & . & . & $\dots$ & . & . & . & . & . \\
%  1 & 0 & 0 & $\dots$ & 11 & 32 & 0 & 5 & 40.9 \\ 
 0 & 1 & 0 & $\dots$ & 10 & 36 & 3 & 2 & 45.81 \\ 
 \hline
\end{tabular}
\label{table:truthtable_confsystem}
%\vspace{-6pt}
\end{center}
\end{table}

% \vspace{-4pt}
\subsection{Uncertainty in the Configurable Software Performance Prediction Problem} \label{sec:uncertainty_perf}

When modeling configurable software performance, there are different sources of uncertainty that can affect the modeling process, such as the measurement bias, the model representation and learning process, and the limited training data \cite{dorn2020mastering}. These sources of uncertainty can be mainly grouped into two types, namely \textit{aleatoric uncertainty} \cite{gurevich2019pairing} and \textit{epistemic uncertainty} \cite{hullermeier2021aleatoric}, which are the two types of uncertainty that our work and recent work \cite{Kendall2017uncertainty, dorn2020mastering} focus on. Aleatoric uncertainty refers to the uncertainty arisen due to the random nature of the system under study whilst epistemic uncertainty refers to the uncertainty occurred due to the differences between the prediction model and the true system under study.

Let us denote a training dataset $\mathcal{D}_{tr} = \{(\textbf{x}_i,y_i)\}_{i=1}^N$ where $y_i = f(\textbf{x}_i)+\epsilon_i$, $\textbf{x}$ is the input data, $f(\textbf{x})$ is the true output value w.r.t. the input $\textbf{x}$, $y$ is the measured output value, and $\epsilon_i$ is the corrupted noise. Suppose that we use a parametric prediction model $f_{\omega}$ (with $\omega$ denoting the model parameters) to model the true function $f$ from the training data $\mathcal{D}_{tr}$. Standard training methods (e.g., gradient descent) generally provide an estimate $\hat{\omega}$ of the parameter, which is the most likely parameter to make the prediction model $f_{\omega}$ to be the same as the true function $f$. However, in practice, there is no guarantee that this model $f_{\hat{\omega}}$ will be the same as the true function $f$ as (i) the representation $f_{\omega}$ might not accurately reflect the true function $f$. Even when the representation $f_{\omega}$ is correct, the estimate $\hat{\omega}$ might not be the same as the true parameter of $f$ as (ii) the training data is finite and most training methods can only provide 100\% accurate estimate when the training data is infinite, and (iii), there is corrupted noise $\epsilon$ in the measured data which makes the estimation process to be not certainly accurate. Here, the issues (i) and (ii) relate to the epistemic uncertainty whilst the issue (iii) relates to the aleatoric uncertainty.

% For configurable software systems, aleatoric uncertainty can be caused by measurement errors which occur when the measurement process is erroneous, or by the representation error caused by the discretization of measurement data for storage and processing \cite{dorn2020mastering}.  For configurable software systems, although epistemic uncertainty could be addressed by collecting the performance measurements of all configurations, each data point still has aleatoric uncertainty ~\cite{dorn2020mastering}. Besides, it is mostly impossible to measure the system performance under all configurations.

% \hy{what problem}

To take into account these uncertainties, a recent approach~\cite{dorn2020mastering} proposed to develop prediction models that not only provide the scalar predictions for the performance, but also provide the confidence intervals of these scalar predictions. The main idea is to perform Bayesian inference over a linear regression model to infer the posterior distribution of the model parameters, and then construct the confidence intervals of the scalar predictions. Although being effective in providing scalar predictions and confidence intervals for small software systems, this approach generally does not perform well on large-scale systems. In this work, we will develop a new method that can provide more accurate scalar performance predictions and more reliable confidence intervals for configurable software systems, especially large-scale systems.

% \vspace{-6pt}
\section{Uncertainty Estimation for Configurable Software Systems via Bayesian Neural Networks}

%\subsection{Overview}

\begin{figure*}
\centerline{\includegraphics[width=172mm,height=78mm]{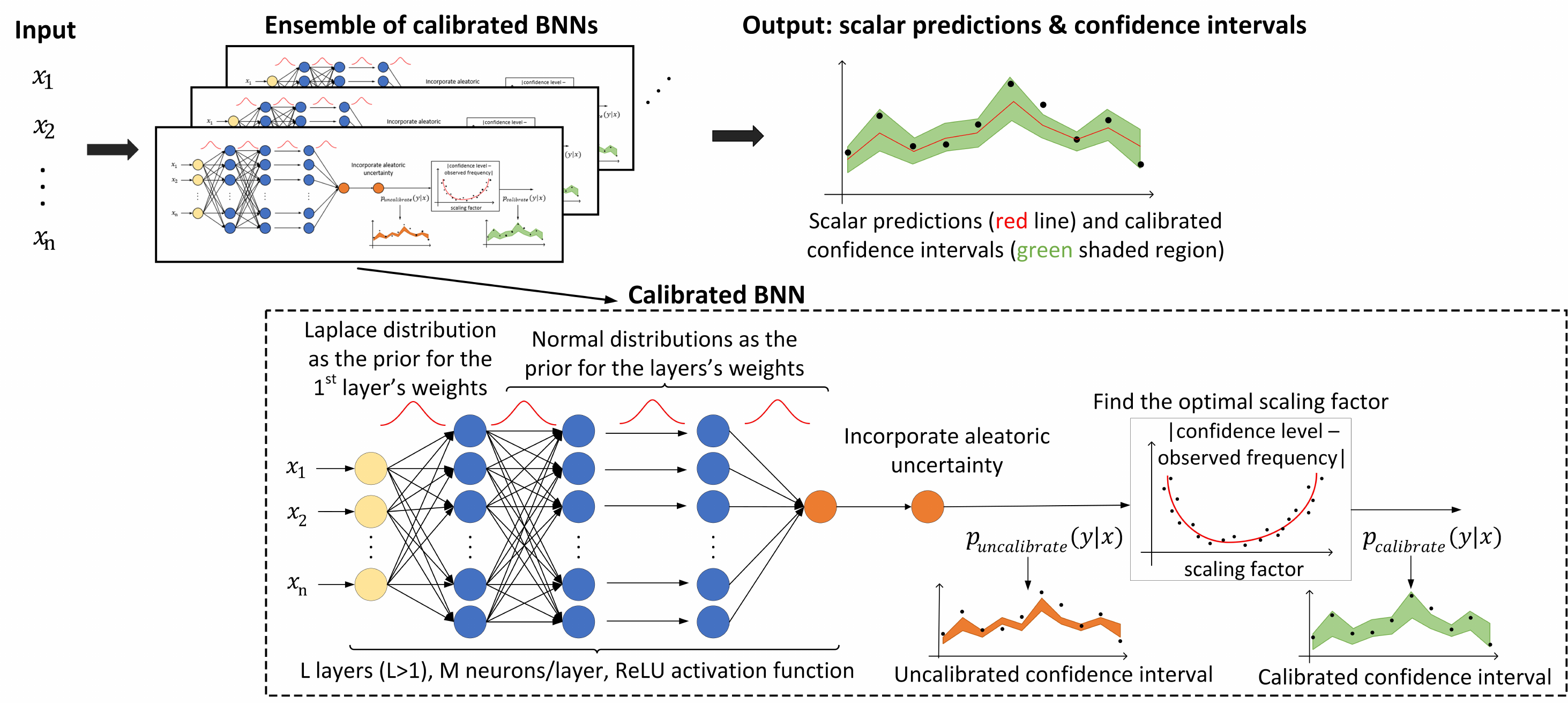}}
%\vspace{-2pt}
\caption{The overall framework of our proposed method, \textbf{BDLPerf}. \textbf{BDLPerf} is an ensemble of calibrated BNNs with each calibrated BNN consisting of two components. The first component is a sparse BNN that can provide both scalar predictions and confidence intervals for configurable software performance. The second component is an uncertainty calibration technique that helps to improve the reliability of the confidence intervals generated by the sparse BNN.}
%\vspace{-4pt}
\label{fig:BDLPerf}
\end{figure*}

An overview of our proposed method, \textbf{BDLPerf}, is presented in Figure \ref{fig:BDLPerf}. \textbf{BDLPerf} is an ensemble of calibrated Bayesian neural networks (BNNs) with each calibrated BNN consisting of two components. The first component is a sparse BNN that can provide both scalar predictions and confidence intervals for configurable software performance (Section \ref{sec:uncertainty_bdl}). The second component is an uncertainty calibration technique that helps to improve the reliability of the confidence intervals generated by the sparse BNN (Section \ref{sec:uncertainty_calibrate}). Finally, a hyper-parameter tuning strategy (Section \ref{sec:hyperpara_tuning}) is applied on the sparse BNN to find its optimal hyperparameters efficiently.

% \vspace{-4pt}
\subsection{Uncertainty Estimation with Bayesian Neural Networks} \label{sec:uncertainty_bdl}

 In this section, we describe in detail how we model the epistemic and aleatoric uncertainty for deep neural networks with Bayesian inference techniques given limited training data.

\subsubsection{Modeling Epistemic Uncertainty with BNNs}

To model the epistemic uncertainty, Bayesian inference techniques aim to generate the posterior probability distribution $p(\omega \vert \mathcal{D}_{tr})$ of the prediction model's parameters, instead of only providing one single estimate $\hat{\omega}$. This parameter probability distribution indicates the probability of a parameter $\omega$ being the true parameter. Given the parameter posterior probability distribution $p(\omega \vert \mathcal{D}_{tr})$, for any new input data $\textbf{x}'$, we can compute the posterior probability distribution of $f(\textbf{x}')$, $p(f(\textbf{x}') \vert \mathcal{D}_{tr})$, as,
\begin{equation}
\begin{aligned}
    \textstyle p(f(\textbf{x}') \vert \mathcal{D}_{tr}) & = \int p(f(\textbf{x}') \vert \omega) p(\omega \vert \mathcal{D}_{tr}) \text{d}\omega \\
    & = \int p(f_{\omega}(\textbf{x}')) p(\omega \vert \mathcal{D}_{tr}) \text{d}\omega,
\end{aligned}
\end{equation}
where $f_{\omega}(\textbf{x}')$ represents the output of the prediction model $f_{\omega}$ w.r.t. the input data $\textbf{x}'$. With the posterior distribution $p(f(\textbf{x}') \vert \mathcal{D}_{tr})$, the scalar prediction of $f(\textbf{x}')$ can be computed as $\mathbb{E}[f(\textbf{x}')] = \int f(\textbf{x}') p(f(\textbf{x}') \vert \mathcal{D}_{tr}) \text{d} f(\textbf{x}')$ where $\mathbb{E}$ denotes the expectation operator. If the posterior probability distribution $p(f(\textbf{x}') \vert \mathcal{D}_{tr})$ is Gaussian, the $\rho$\% confidence interval ($0 \leq \rho \leq 100$) of $f(\textbf{x}')$ can be computed as $\mathbb{E}[f(\textbf{x}')] \pm z(\rho) \mathbb{SD}[f(\textbf{x}')]$ where $\mathbb{SD}$ denotes the standard deviation operator, i.e., $\mathbb{SD}[f(\textbf{x}')] = \sqrt{\int (f(\textbf{x}') - \mathbb{E}[f(\textbf{x}')])^2 p(f(\textbf{x}') \vert \mathcal{D}_{tr}) \text{d} f(\textbf{x}')}$, and $z(\rho)$ denotes the z-score of the confidence level $\rho$. Bayesian inference techniques have been shown to be effective in generating posterior probability distribution of the prediction model's parameters, and thus effective in generating the scalar predictions and confidence intervals (i.e., uncertainty estimates) for the outputs \cite{Gal2016Uncertainty, Gal2016MCdropout, Lakshminarayanan2017, AbdarUncertainty2021, HullermeierUncertainty2021}.

In general, Bayesian inference techniques could be applied to different parametric prediction models including neural network and linear regression. There are a number of well-known Bayesian inference techniques such as the Variational Inference (VI) method \cite{Jordan1999VI}, Markov Chain Monte Carlo (MCMC) \cite{Duane1987MCMC}, Monte Carlo Dropout \cite{Gal2016MCdropout}, Ensemble \cite{Lakshminarayanan17}. Whilst the VI and MCMC techniques are traditional Bayesian inference techniques that can be applied to any parametric prediction models, the Monte Carlo Dropout and Ensemble techniques are generally applied to deep neural networks. When the prediction model is a neural network, then the network obtained after the Bayesian inference process is called a Bayesian neural network (BNN) \cite{Mackay1992, Neal1995, Gal2016Uncertainty}.

In this work, we use the feedforward neural network (FNN) as the architecture of the prediction model as it has been shown that FNNs can accurately represent complex functions \cite{Goodfellow2016DL, Barron1993, Hornik1989, Funahashi1989} including the performance of configurable software systems \cite{ha2019deepperf}. To perform Bayesian inference on the FNN to obtain a BNN, we employ the VI method \cite{Jordan1999VI} due to its effectiveness \cite{Gal2016Uncertainty, Abdar2021UncertaintyRv}. The key idea of the VI method is to consider the Bayesian inference problem as an optimization problem, and solve this optimization problem over a family of tractable distributions to find a distribution that is closest to the true posterior distribution $p(\omega \vert \mathcal{D}_{tr})$. For the configurable software performance prediction problem, as the training data is limited, the accuracy of the BNNs obtained by the VI method may not be optimal. We therefore propose to use the Ensemble method to generate and combine multiples BNNs together as the Ensemble method is well known to be able to enhance individual model's accuracy \cite{Lakshminarayanan17}. To construct the ensemble, we divide the training data into $K$ folds ($K>1$), and train a BNN using data in $K-1$ folds with the VI method. We thus obtain an ensemble of $K$ BNNs trained on different data subsets of the training data.

Finally, it has been widely recognized that for a configurable software system, its performance depends only on a limited number of configuration options and their interactions \cite{Jamshidi2017transferlearningperf,siegmund2015performance, Siegmund2012performanceauto}. Previous approaches \cite{ha2019deepperf, dorn2020mastering, ha2019performance} have incorporated this knowledge into the modeling process by making the prediction model's parameters to be sparse. In this work, we also aim to train sparse BNNs. Similar to \cite{ha2019deepperf}, for each BNN in the ensemble, we make the parameters of the BNN's first layer sparse by specifying the prior distribution of this layer's parameters to be a Laplace distribution. It has been shown that specifying a Laplace prior distribution over the layer's parameters is equivalent to apply the L1 regularization to the layer \cite{Tibshirani1996lasso}, and thus, resulting in a sparse layer. To control the sparsity level of the BNN, we set the scale parameter of the Laplace distribution as a hyperparameter to be tuned.

Now, we have constructed an ensemble of sparse BNNs that can incorporate the epistemic uncertainty. In the next section, we describe how to model the aleatoric uncertainty.

% \vspace{6pt}
\subsubsection{Modeling Aleatoric Uncertainty}
As mentioned in Section \ref{sec:uncertainty_perf}, for the configurable software performance prediction problem, the aleatoric uncertainty is generally caused by measurement errors, e.g., corrupted noise or erroneous measurement process. There are generally two approaches to model the aleatoric uncertainty \cite{Kendall2017uncertainty, dorn2020mastering}. The first approach is to use a homoscedatic model which assumes all the data points have the same noise variance $\sigma$. The second approach is to use a heteroscedastic model which assumes each data point $\textbf{x}$ has its own noise variance $\sigma(\textbf{x})$. In this work, we use a heteroscedastic model as it is more accurate. For each BNN in the ensemble, to incorporate the aleatoric uncertainty via the heteroscedastic model, we formulate a Normal distribution around the output of the BNN,
\begin{equation} \label{eq:aleatoric-uncertainty}
    f_{\omega,\text{final}}(\textbf{x}) = \mathcal{N}(f_{\omega}(\textbf{x}), \sigma (\textbf{x})),
\end{equation}
where $f_{\omega,\text{final}}(\textbf{x})$ denotes the final output of the BNN w.r.t. the input $\textbf{x}$, $f_{\omega}(\textbf{x})$ denotes the output of the original BNN when only epistemic uncertainty is incorporated, and $\sigma (\textbf{x})$ denotes the variance of the measurement errors. Eq. (\ref{eq:aleatoric-uncertainty}) shows that the final output of each BNN in the ensemble now takes into account the possible measurement errors during the data measurement process, and thus includes both aleatoric and epistemic uncertainty. The noise variance function $\sigma()$ will be learned along with the BNN's parameters.

\subsection{Effective Uncertainty Calibration Strategy} \label{sec:uncertainty_calibrate}

% Finally, an important quantification when evaluating the quality of the uncertainty estimates is their \textit{calibration} property. \textit{Calibration} refers to the concept which measures the discrepancy between the subjective forecasts and the (empirical) long-run frequencies \cite{Lakshminarayanan17, Dawid1982calibration}. A method that can generate well-calibrated uncertainty estimates means that for all $\rho, 0\leq \rho \leq 100$, the $\rho\%$-confidence intervals contain the ground truth values $\rho\%$ of the time. For example, in Figure \ref{fig:uncertainty_illustration}, the uncertainty estimates for the 95\% percentile are well-calibrated as there are 95\% of the data points within the 95\% confidence intervals. When this condition is not satisfied, we say that the uncertainty estimates are mis-calibrated. 

Bayesian inference techniques can incorporate uncertainty and generate confidence intervals for a prediction model, however, their generated confidence intervals are generally not well-calibrated \cite{Kuleshov2018}. We call a generated $\rho$\% confidence interval to be well-calibrated (or reliable) if it contains the ground truths $\rho$\% times \cite{Kuleshov2018, dorn2020mastering}. For example, in Figure \ref{fig:uncertainty_illustration}, the generated 95\% confidence interval is well-calibrated as there are 95\% of the ground truths within the confidence interval. A method provides well-calibrated confidence intervals if for all confidence level $\rho (0 \leq \rho \leq 100)$, the generated $\rho$\% confidence interval is well-calibrated. In the sequel, we will denote $\alpha_{\rho}$ as the percent of the ground truths to be in the $\rho$\% confidence interval; $\alpha_{\rho}$ is usually called the \textit{observed frequency} w.r.t. $\rho$. A $\rho$\% confidence interval is well-calibrated when $\alpha_{\rho} = \rho$.

There are a number of methods developed to calibrate the confidence intervals generated by a Bayesian inference technique \cite{Guo2017calibration}. In this work, we take inspiration from one of these calibration techniques, the Platt scaling technique \cite{Platt99scaling, Guo2017calibration}, and develop it to work with the limited training data scenario as in the configurable software performance problem. We choose the Platt scaling technique as it is simple yet effective. It is also model-diagnostic, i.e., it is general and can be applied to various types of prediction model.

The main idea of Platt scaling is to split the training data into two parts: train the prediction model with data in one part, evaluate the generated confidence intervals with data in the other part, then based on the evaluation data, find a new confidence level $\rho_{\text{cal}}$ so that the confidence interval w.r.t this new confidence level will contain the ground truths $\rho$\% of time. At testing time, we can generate this new confidence interval instead of the $\rho$\% confidence interval. However, the Platt scaling technique generally does not work well for the configurable software performance prediction problem as the training data in this case is limited. The first issue is that when a prediction model is trained with a limited training data, its generated confidence intervals can be poorly-calibrated. For various values of $\rho$, the generated $\rho$\% confidence interval might be too small that it may not contain any ground truth, making it impossible to find $\rho_{\text{cal}}$. The second issue is that even when the generated $\rho$\% confidence interval contains the ground truths, the evaluation data consists of very limited data and might not be enough to represent the whole data population, hence, the found value $\rho_{\text{cal}}$ might not be accurate.

To address the first issue, for each confidence level $\rho$, we aim to find a scaling factor $\zeta_{\rho}$ such that when scaling the generated $\rho$\% confidence interval with this scaling factor, the new confidence interval will contain the ground truths $\rho$\% times. To find this scaling factor, we train the prediction model using one part of the training data, generate the $\rho$\% confidence interval on the other part, and then search for the value $\zeta_{\rho}$ that minimizes the difference between $\rho$ and the observed frequency $\alpha_{\rho}$ obtained when the generated $\rho$\% confidence interval is scaled with $\zeta_{\rho}$. At testing time, we scale the generated $\rho$\% confidence interval with this scaling factor. By the definition of $\zeta_{\rho}$, the new confidence interval will be well-calibrated as it contains the ground truths $\rho$\% times.

To address the issue of possible inaccurate computation of the scaling factor $\zeta_{\rho}$ due to limited evaluation data, we make use of our ensemble approach. Each BNN in the ensemble is trained using data in $K-1$ folds, and the scaling factor w.r.t. to each BNN is searched using data in the remaining fold. At testing time, the final confidence interval is combined from the adjusted confidence intervals of all the $K$ BNNs in the ensemble. This helps to mitigate issues caused by an abnormal data splitting and ensure a more accurate calibration process.

\subsection{Efficient Hyperparameter Tuning} \label{sec:hyperpara_tuning}

 Training a BNN involves the process of tuning various hyperparameters which are critical to ensure high model's accuracy. As the BNN training process could be expensive, the challenge is to obtain an optimal set of hyperparameters with a minimal number of training evaluations. In this section, we develop a new technique to identify the optimal hyperparameter set efficiently. Our hyperparameter tuning technique combines Bayesian Optimization (BO) \cite{Snoek2012} and our developed search space identification strategy. BO is a powerful optimization method to find the global optimum of an unknown objective function by sequential queries \cite{Jones1998, Mockus1978, Snoek2012}. It has been shown to be effective and efficient in finding the global optimum of black-box functions, especially in tuning hyperparameters of deep neural networks \cite{Shahriari16a, Snoek2012, Turner2021BO}.
 
%  BO is a powerful optimization method to find the global optimum of an unknown objective function by sequential queries ~\cite{Jones1998, Mockus1978, Shahriari16a, Snoek2012}. Its main idea is to use a surrogate model to approximate the objective function from the current evaluation data, and then construct an acquisition function to select the next most informative data point to be evaluated. The process is conducted iteratively until the evaluation budget is depleted and the optimum is selected from the evaluation data. BO has been shown to be effective and efficient in finding the global optimum of black-box functions, especially in tuning hyperparameters of deep neural networks \cite{Shahriari16a, Snoek2012, Turner2021BO}.

Any search method (including BO) requires a search space to search for the optimal hyperparameters. Setting a too small search space will result in not finding the most optimal hyperparameter set whilst setting a too large search space will result in high computation time. To address this problem, we propose an efficient strategy to identify an optimal search space based on the nature of configurable software performance. When using FNN to model a complex function, the most important hyperparameter is generally the number of layers as it defines the network's complexity, thus defines its accuracy. Therefore, we aim to find the search space of the number of layers whilst fixing the search spaces of other hyperparameters to be some common choices. Our strategy is as follows. We first split the training data into two parts, one part for training the prediction model and one part for evaluating the model accuracy. We start with the number of layers being 1, use BO to tune the remaining hyperparameters, and evaluate the BNN's accuracy. We then increase the number of layers by 1, perform as previously to find the BNN's accuracy w.r.t. the new number of layer. This process is conducted repeatedly, until we find the number of layers whose BNN's accuracy is lower than the accuracy of the previous number of layers. The process is then terminated and the chosen number of layers is the one with the highest model accuracy. Finally, we use BO to tune all remaining hyperparameters of this chosen number of layers. Note, in the final BO process, we use transfer learning to include all previous evaluations so that BO can find the optimal hyperparameters with a minimal number of evaluations.

% We first split the training data into two parts: train and validation. We start with the number of layer being 1, use BO to tune the remaining hyperparameters, and evaluate the BNN's accuracy on the validation data. We then increase the number of layers by 1, perform as previously to find the BNN's accuracy w.r.t. the new number of layer. This process is conducted repeatedly, until we find the number of layer whose BNN's accuracy is lower than the accuracy of the previous layer. The optimal number of layer is the number of layer with the highest BNN's accuracy. Finally, we use BO to tune all the remaining hyperparameters of this optimal number of layer. Note, in the final BO process, we use transfer learning to include all previous evaluations so that BO can find the optimal hyperparameters with a minimal number of evaluations.

% the optimal number of layer first, then tune the remaining hyperparameters: the number of epoch, the learning rate, and the scaling factor of the Laplace prior distribution. Since 

% \vspace{-2pt}
\subsection{Model Training  and Tool Implementation}

We implement our proposed technique, \textbf{BDLPerf}, using Python 3.9. The BNNs are implemented with TensorFlow Probability 0.15.0 \cite{TFProbability}. We perform some data preprocessing steps as in \cite{dorn2020mastering} and \cite{ha2019deepperf}. Specifically, we use the entropy method developed in \cite{dorn2020mastering} to remove the multicollinearity issue in the whole dataset. We then normalize the output (i.e., performance values) between 0 and 100 as in \cite{ha2019deepperf} so that the parameters of the BNNs will not be too small.

% \hy{is this a pre-processing step of the proposed approach? did you do it for the test set as well}

We train the BNNs using the Adam optimizer \cite{Kingma2015Adam} with a scheduled learning rate that when the epoch is larger than $1000$, the learning rate is decreased gradually with an exponential decay rate of -0.001. The hyperparameters of the Adam optimizer are set using the default values of TensorFlow Probability 0.15.0. We set the loss function of the VI method to be the Kullback-Leibler divergence \cite{Kullback1951KLdivergence} whilst the loss function to train the BNNs is the mean square error. The number of samples for the Bayesian inference is 300. For the ensemble, we set the number of folds, $K$, to be 3 with the number of data points in each fold is equal.

For our proposed uncertainty calibration technique, we use grid search to search for the optimal scaling factor. The search range is set from $[0.01 \zeta_{\max}, 10\zeta_{\max}]$ where $\zeta_{\max}$ is the maximum scaling factor that makes all the ground truths in the evaluation dataset (one fold in the training data) belong to the confidence interval scaled by this value. 

% \hh{Do we need to include these hyperparameter settings (in the comments below)? As they are important to understand about our proposed hyperparameter tuning strategy?}\hy{not sure if we have space for these details, maybe can describe it in online appendix/webpage if space is limited. (say detailed can be found at ...)}

For our hyperparameter tuning strategy, we use the BO implementation of the package GPyOpt \cite{gpyopt2016}. The hyperparameters of BO are set using the default values of GPyOpt. In each layer, we tune the number of epochs, the learning rate, the number of neurons/layer, and the scale parameter of the Laplace distribution. We initialize BO with 4 random data points and then conduct the BO process with 12 iterations. When conducting the final BO process to search for other hyperparameters of the optimal layer, we use all the previous evaluations as the initial data and then conduct the BO process with 8 iterations. The search space of the hyperparameters is set as follows. The learning rate domain is from $10^{-4}$ to $0.1$ whilst the scale parameter of the Laplace distribution is from $10^{-4}$ to $1$. These are the common search spaces for these two hyperparameters. The number of epoch is within the discrete set $\{500, 1000, 2000\}$, as inspired from \cite{ha2019deepperf}. The number of neurons/layer is within the set $\{n, 2n, 4n\}$ with $n$ being the number of configuration options. This is to ensure that the BNN is more complex for larger configurable software systems. Besides, note that, as the number of layers controls the complexity of the BNN and we have a procedure to find the optimal number of layers, so the number of neurons/layer can generally be set as any values. Finally, for the cross validation to generate the ensemble BNN, we split the data into 1/3 of training and 2/3 of validation. The rationale behind the idea of having more validation data than training data in the uncertainty calibration process is that we want to compute more accurately the observed frequencies, thus generate the right calibrated confidence intervals. 

% \hy{can say some more about the BDL settings/parameters}
% \hy{say why do you set the values like these?}

\section{Experimental Setup} \label{sec:exp-setup}

% To evaluate the efficacy of our proposed method \textbf{BDLPerf}, we conduct an extensive set of experiments on a number of real-world configurable software systems. In this section, we describe in details our experiment setup, particularly, the research questions we aim to answer (Section \ref{sec:rquestion}), the subject systems (Section \ref{sec:systems}) and the baseline methods (Section \ref{sec:baselines}) used in our evaluation.

\subsection{Research Questions} \label{sec:rquestion}

To evaluate the efficacy of our proposed method \textbf{BDLPerf}, we aim to answer the following research questions (RQs):

\vspace{0.17cm}
\noindent
\fbox{\begin{minipage}{24.3em}
\textbf{RQ1:} Can \textbf{BDLPerf} accurately predict software performance values as scalar predictions?
\end{minipage}}
\vspace{0.01cm}

\noindent
This RQ is to demonstrate that even though \textbf{BDLPerf} also provides the confidence intervals, it still guarantees to provide accurate scalar predictions. To answer this RQ, we will compare the accuracy of the scalar predictions by \textbf{BDLPerf} and the baseline methods.

\vspace{0.25cm}
\noindent
\fbox{\begin{minipage}{24.3em}
\textbf{RQ2:} Can \textbf{BDLPerf} provide accurate and reliable confidence intervals for the scalar predictions?
\end{minipage}}
\vspace{0.01cm}

\noindent
This RQ is to evaluate the accuracy and reliability of the confidence intervals generated by \textbf{BDLPerf}. To answer this RQ, we will compare \textbf{BDLPerf} with the baseline methods in generating confidence intervals.

\vspace{0.25cm}
\noindent
\fbox{\begin{minipage}{24.3em}
\textbf{RQ3:} Is it necessary to develop an uncertainty calibration method for the configurable software performance?
\end{minipage}}
\vspace{0.01cm}

\noindent
This RQ is to show that in the limited training data scenario as in the configurable software performance prediction problem, an uncertainty calibration technique is needed. To answer this RQ, we will compare the confidence intervals generated by \textbf{BDLPerf} and those generated by methods without uncertainty calibration or with standard uncertainty calibration.

\vspace{0.24cm}
\noindent
\fbox{\begin{minipage}{24.3em}
\textbf{RQ4:} What is the time cost of \textbf{BDLPerf} in providing scalar predictions and confidence intervals?
\end{minipage}}
\vspace{0.01cm}

\noindent
This RQ is to evaluate the time cost of our proposed method. It aims to show that even though \textbf{BDLPerf} is a deep learning based method, its running time is reasonable.

The detailed setup and evaluation metrics for each RQ will be clearly described in Section \ref{sec:exp-results}. In general, we use various benchmark metrics on scalar prediction and uncertainty estimation to compare the performance of the methods.

\subsection{Subject Systems} \label{sec:systems}

In our evaluation, we use 10 real-world configurable software systems, as presented in Table \ref{table:systems}. These systems are common benchmarks that are widely used to evaluate methods for the software performance prediction problem \cite{ha2019deepperf, dorn2020mastering, Kaltenecker2020samplingdistance}. The systems have different characteristics and are from different application domains including database management system (BDB-C), compiler (LLVM), multi-grid solver (Dune), file archive utility (7z, lrzip), image processing framework (Hipacc), video encoders (x264, VP9), code optimizer (Polly) and garbage collector (JavaGC). The performance measured are the response time, solving time, compilation time, run time, encoding time, compression time, or the energy consumption. The number of configuration options of these systems ranges from 11 to 54 including both binary and/or numeric configuration options. The number of valid configurations ranges from 400 to approximately 200000. These systems' performance values were measured carefully by Kaltenecker et al. \cite{Kaltenecker2020samplingdistance} with various guaranteeing techniques so as to ensure their correctness. The measurements took multiple years of CPU time in total. The detailed description of these systems and how they were measured can be found in \cite{dorn2020mastering, Kaltenecker2020samplingdistance}. Table ~\ref{table:systems} summarizes some notable properties of these subject systems.

\begin{table}
\caption{Overview of the subject systems with domain, number of valid configurations $\vert \mathcal{C} \vert$, number of configuration options $\vert \mathcal{O} \vert$, and the kind of performance for prediction.}
\centering
\label{table:systems}
  \begin{tabular}{m{25pt}<{\centering} m{58pt}<{\centering}  m{20pt}<{\centering}  m{8pt}<{\centering}    m{68pt}<{\centering}  }\hline  
       System   & Domain & $\vert \mathcal{C} \vert$ &  $\vert \mathcal{O} \vert$ &  Performance \\ \hline 

7z & File archive utility & 68640 & 44 & Compression time \\
BDB-C & Embedded database & 2560 & 18 & Response time \\
Dune & Multigrid solver & 2304 & 32 & Solving time \\
Hipacc & Image processing & 13485 & 54 & Solving time \\
LLVM & Compiler infrastructure & 1024 & 11 & Compilation time \\
lrzip & File archive utility & 432 & 19 & Compression time \\
Polly & Code optimizer & 60000 & 40 & Runtime \\
x264 & Video encoder & 1152 & 16 & Enconding time Engery consumption \\
VP9 & Video encoder & 216000 & 42 & Enconding time Engery consumption \\
JavaGC & Garbage collector & 193536 & 39 & Time \\
 \hline
  \end{tabular}
\end{table}

% \begin{table}
% \caption{Overview of the subject systems with domain, number of binary configuration options \#Binary, number of numeric configuration options \#Numeric, and number
% of valid configurations \#Configs.}
% \label{table:systems}
%   \begin{tabular}{m{30pt}<{\centering} m{60pt}<{\centering}  m{30pt}<{\centering}  m{30pt}<{\centering}    m{30pt}<{\centering}  }\hline  
%       System   & Domain & \#Binary &  \#Numeric &  \#Configs \\ \hline 

% 7z & File archive utility & 44 & 0 & 68640 \\
% BDB-C & Embedded database & 18 & 0 & 2560 \\
% Dune & Multigrid solver & 8 & 3 & 2304 \\
% HIPAcc & Image processing & 31 & 2 & 13485 \\
% LLVM & Compiler infrastructure & 11 & 0 & 1024 \\
% lrzip & File archive utility & 19 & 0 & 432 \\
% Polly & Code optimizer & 40 & 0 & 60000 \\
% x264 & Video encoder & 16 & 0 & 1152 \\

%  \hline
%   \end{tabular}
% \end{table}

\subsection{Baselines} \label{sec:baselines}
% \hy{describe the baselines to be compared with}

We compare our proposed method, \textbf{BDLPerf}, with the state-of-the-art method \textbf{P4} proposed in \cite{dorn2020mastering}. To the best of our knowledge, this is the only method that can provide confidence intervals along with scalar predictions for configurable software performance. The main idea of \textbf{P4} is to represent the performance of configurable software system using a Lasso regression model. Then probabilistic programming is used to perform Bayesian inference so as to obtain the posterior probability distribution of the model's parameters, and thus posterior probability distribution of the performance values. \textbf{P4} has two settings, namely, $\Pi_{ho}$ and $\Pi_{he}$. The setting $\Pi_{ho}$ is based on a homoscedastic aleatoric uncertainty model whilst the setting $\Pi_{he}$ is based on a heteroscedastic aleatoric uncertainty model. In this work, we will compare our proposed method \textbf{BDLPerf} with both settings of \textbf{P4}, $\Pi_{ho}$ and $\Pi_{he}$.  To replicate the results of these two settings, we use the code published on their Github page \cite{P4Uncertainty}.

% We ran the two variants with their default settings.

\section{Experimental Results} \label{sec:exp-results}

% In this section, we show and discuss the experimental results for the four research questions described in Section \ref{sec:rquestion}.

\subsection{RQ1: Accuracy of the Scalar Predictions of \textbf{BDLPerf}}

\begin{table*}[htbp]
  \centering
  \caption{Experimental results showing the mean and margin values of the MAPE scores of our proposed method \textbf{BDLPerf} and the baseline method \textbf{P4} \cite{dorn2020mastering} with two settings $\Pi_{ho}$, $\Pi_{he}$. The lower the MAPE scores, the better the methods. Experiments are repeated 20 times. In the column \textit{Better Algorithm}, we report the better method using t-test with p-value of 0.05.}
  
    \begin{tabular}{m{29pt}<{\centering} | m{24pt}<{\centering} | m{19pt}<{\centering}  m{24pt}<{\centering} | m{19pt}<{\centering}  m{24pt}<{\centering} | m{41pt}<{\centering} || m{19pt}<{\centering}  m{24pt}<{\centering} | m{19pt}<{\centering}  m{24pt}<{\centering} | m{41pt}<{\centering} }\hline 
    
      \multirow{2}{*}{System} & \multirow{2}{*}{t-wise} & \multicolumn{2}{c}{BDLPerf} &  \multicolumn{2}{c}{P4($\Pi_{ho}$)} & \multirow{2}{39pt}{\centering Better Algorithm} & \multicolumn{2}{c}{BDLPerf} & \multicolumn{2}{c}{P4($\Pi_{he}$)} & \multirow{2}{39pt}{\centering Better Algorithm}  \\
       &  & Mean & Margin & Mean & Margin & & Mean & Margin & Mean & Margin &  \\ \hline
     \multirow{3}{*}{7z} & 1 & \textbf{53.2} & 1.8 & 65.0 & 9.4e-2 & BDLPerf & \textbf{53.2} & 1.8 & 61.2 & 1.2e-1 & BDLPerf \\ 
                         & 2 & \textbf{15.7} & 0.5 & 73.1 & 7.1e-2 & BDLPerf & \textbf{15.7} & 0.5 & 65.5 & 4.8e-2 & BDLPerf \\ 
                         & 3 & \textbf{9.2} & 0.3 & 14.5 & 3.0e-2 & BDLPerf & \textbf{9.2} & 0.3 & \textbf{9.2} & 1.5e-3 & Same \\ \hline
  \multirow{3}{*}{BDB-C} & 1 & \textbf{63.4} & 4.0 & 122.9 & 8.0e-2 & BDLPerf & \textbf{63.4} & 4.0 &  127.9 & 6.2 & BDLPerf \\ 
                         & 2 & \textbf{22.3} & 3.7 & 51.5 & 8.7e-2 & BDLPerf & \textbf{22.3} & 3.7 & 59.3 & 5.4 & BDLPerf \\ 
                         & 3 & \textbf{7.7} & 1.6 & 16.1 & 5.2e-2 & BDLPerf & \textbf{7.7} & 1.6 & 58.1 & 12.3 & BDLPerf \\ \hline
   \multirow{3}{*}{Dune} & 1 & 18.8 & 0.7 & \textbf{17.1} & 5.7e-3 & P4 & 18.8 & 0.7 & \textbf{17.1} & 4.0e-3 & P4 \\ 
                         & 2 & \textbf{10.9} & 0.3 & 13.0 & 1.7e-2 & BDLPerf & \textbf{10.9} & 0.3 & 12.6 & 6.6e-3 & BDLPerf \\ 
                         & 3 & \textbf{4.9} & 0.3 & 8.1 & 1.1e-2 & BDLPerf & \textbf{4.9} & 0.3 & 7.6 & 3.4e-3 & BDLPerf \\ \hline
 \multirow{3}{*}{Hipacc} & 1 & \textbf{24.4} & 1.9 & 53.0 & 3.0e-2 & BDLPerf & \textbf{24.4} & 1.9 & 51.9 & 4.9e-2 & BDLPerf \\ 
                         & 2 & \textbf{12.6} & 1.1 & 17.6 & 1.1e-2 & BDLPerf & \textbf{12.6} & 1.1 & 16.9 & 7.3e-3 & BDLPerf \\ 
                         & 3 & \textbf{2.4} & 0.1 & 8.9 & 3.3e-3 & BDLPerf & \textbf{2.4} & 0.1 & 7.9 & 9.8e-4 & BDLPerf \\ \hline
   \multirow{3}{*}{LLVM} & 1 & 9.0 & 0.9 & \textbf{6.8} & 6.4e-3 & P4 & 9.0 & 0.9 & \textbf{6.9} & 6.5e-3 & P4 \\ 
                         & 2 & \textbf{5.0} & 0.8 & 5.7 & 2.9e-3 & BDLPerf & \textbf{5.0} & 0.8 & 5.8 & 3.6e-3 & BDLPerf \\ 
                         & 3 & 3.0 & 0.4 & \textbf{2.7} & 1.6e-3 & P4 & 3.0 & 0.4 & \textbf{2.8} & 1.4e-3 & Same \\ \hline
  \multirow{3}{*}{Polly} & 1 & 48.0 & 0.9 & \textbf{30.9} & 6.3e-3 & P4 & 48.0 & 0.9 & \textbf{31.3} & 4.0e-3 & P4 \\ 
                         & 2 & \textbf{8.1} & 1.2 & 11.4 & 4.0e-3 & BDLPerf & \textbf{8.1} & 1.2 & 12.4 & 6.7e-3 & BDLPerf \\ 
                         & 3 & \textbf{3.1} & 0.3 & 11.1 & 1.9e-3 & BDLPerf & \textbf{3.1} & 0.3 & 11.0 & 1.1e-3 & BDLPerf \\ \hline
  \multirow{3}{*}{lrzip} & 1 & 52.8 & 3.2 & \textbf{33.8} & 1.4e-1 & P4 & 52.8 & 3.2 & \textbf{32.7} & 2.6e-2 & P4 \\ 
                         & 2 & \textbf{29.9} & 13.3 & 69.4 & 6.7e-1 & BDLPerf & \textbf{29.9} & 13.3 & 54.0 & 9.8e-2 & BDLPerf \\ 
                         & 3 & \textbf{7.0} & 2.9 & 36.0 & 6.9e-1 & BDLPerf & 7.0 & 2.9 & \textbf{5.0} & 1.5e-2 & Same \\ \hline
   \multirow{3}{*}{x264} & 1 & 15.9 & 3.6 & \textbf{9.5} & 2.4e-2 & P4 & 15.9 & 3.6 & \textbf{7.7} & 1.3e-2 & P4 \\ 
                         & 2 & \textbf{5.8} & 2.9 & 16.3 & 3.1e-2 & BDLPerf & \textbf{5.8} & 2.9 & 9.4 & 1.1e-2 & BDLPerf \\ 
                         & 3 & \textbf{1.3} & 0.2 & 3.8 & 1.4e-2 & BDLPerf & \textbf{1.3} & 0.2 & 1.4 & 9.9e-4 & Same \\ \hline
    \multirow{3}{*}{VP9} & 1 & \textbf{108.4} & 10.5 & 148.9 & 0.3 & BDLPerf & \textbf{108.4} & 10.5 & 164.3 & 6.9 & BDLPerf \\ 
                         & 2 & \textbf{28.7} & 1.3 & 106.7 & 0.1 & BDLPerf & \textbf{28.7} & 1.3 & 135.6 & 14.3 & BDLPerf \\ 
                         & 3 & \textbf{11.7} & 0.7 & 81.9 & 0.1 & BDLPerf & \textbf{11.7} & 0.7 & 112.7 & 12.0 & BDLPerf \\ \hline
 \multirow{3}{*}{JavaGC} & 1 & 42.6 & 2.4 & \textbf{40.9} & 5e-3 & Same & 42.6 & 2.4 &  \textbf{40.8} & 5e-3 & Same \\ 
                         & 2 & \textbf{32.9} & 2.5 & 63.9 & 4e-2 & BDLPerf & \textbf{32.9} & 2.5 & 51.5 & 7e-2 & BDLPerf \\ 
                         & 3 & \textbf{10.5} & 0.9 & 31.4 & 2e-2 & BDLPerf & \textbf{10.5} & 0.9 & 14.2 & 3e-3 & BDLPerf \\ \hline
    \end{tabular}%
  \label{table:rq1}%
\end{table*}%

In this RQ, we compare the effectiveness of our method, \textbf{BDLPerf}, with the baseline method \textbf{P4} on two settings $\Pi_{ho}$ and $\Pi_{he}$ in generating scalar predictions for configurable software performance.

\paragraph{Setup} We use the same experiment setup  as in \cite{dorn2020mastering}. For each subject system, we use three different sampling strategies, in particular, the t-wise sampling with $t \in \{ 1,2,3\}$, to generate the training datasets. All the measurements of each system are used as the testing dataset. We use the prediction model trained by each method to predict the scalar performance of all the configurations in the testing dataset. We then use the Mean Absolute Percentage Error (MAPE) score to quantify the prediction accuracy of the models. The MAPE score is computed as,
\begin{equation}
    \text{MAPE}(C) = \dfrac{1}{\vert C \vert} \sum\nolimits_{c \in C} \dfrac{\vert \text{predicted}_c - \text{true}_c \vert}{\text{true}_c} \times 100\%,
\end{equation}
where $c$ denotes a configuration, $C$ denotes the testing dataset, $\text{predicted}_c$ denotes the scalar prediction for configuration $c$, and $\text{true}_c$ denotes the true performance value of configuration $c$. We use this metric as it is widely used to evaluate the accuracy of scalar predictions, especially for configurable software performance \cite{siegmund2015performance, Guo2018Decart, ha2019deepperf, dorn2020mastering}. To ensure the stability and consistency of the evaluation, we repeat the experiments 20 times. We then report the mean and the 95\% confidence interval (margin) of the MAPE scores obtained after 20 experiments. Finally, we use t-test with the significant level of 0.05 to statistically compare the MAPE scores between \textbf{BDLPerf} and \textbf{P4}. Since \textbf{P4} has two settings ($\Pi_{ho}$ and $\Pi_{he}$), we sequentially compare \textbf{BDLPerf} with \textbf{P4} in each setting.

% \hy{we perform t-test pairwise?}
% \hy{not clear, could be  "better than the other two methods"? }
% \vspace{-0.1cm}
\paragraph{Results} In Table \ref{table:rq1}, we report the MAPE scores of \textbf{BDLPerf} and \textbf{P4} with two settings $\Pi_{ho}$ and $\Pi_{he}$ on 10 subject systems with different t-wise sampling strategies. Note that the lower the MAPE scores, the higher the prediction accuracy (i.e., the better the methods). For the setting $\Pi_{ho}$, we can see that \textbf{BDLPerf} statistically outperforms \textbf{P4} on 7z, BDB-C, Hipacc, and VP9 for all sampling strategies and outperforms other systems on majority of sampling strategies. This shows the accuracy of \textbf{BDLPerf} is much higher compared to \textbf{P4}. In total, \textbf{BDLPerf} statistically outperforms \textbf{P4} with the setting $\Pi_{ho}$ on 23/30 cases and performs similarly on 1/30 cases. On the other hand, with the setting $\Pi_{he}$, \textbf{BDLPerf} also statistically outperforms \textbf{P4} on 20/30 cases and performs similarly on 5/30 cases. Finally, it is also worth noting that \textbf{BDLPerf} especially outperforms \textbf{P4} on large subject systems (e.g., 7z, Hipacc, Polly, VP9, JavaGC) by a very large margin. For example, for VP9, the MAPE scores of \textbf{BDLPerf} are 108.4\%, 28.7\%, and 11.7\% whilst the corresponding best MAPE scores of P4 are 148.9\%, 106.7\%, and 81.9\%, which are much worse compared to \textbf{BDLPerf}. This demonstrates the effectiveness of using deep neural networks as the prediction models for the performance of large configurable software systems.

\subsection{RQ2: Quality of Confidence Intervals by \textbf{BDLPerf}}

In this RQ, we compare the quality of the confidence intervals generated by our proposed method \textbf{BDLPerf} and the baseline method \textbf{P4} on two settings $\Pi_{ho}$ and $\Pi_{he}$.

\paragraph{Setup} To generate the training and testing datasets, we use the same experiment setup as in RQ1. To measure the quality of the generated confidence intervals of all the methods, we use the calibration metric, $cal$ \cite{Kuleshov2018}, which is the traditional metric for evaluating the accuracy and reliability of the generated confidence intervals. The main idea of the $cal$ metric is to assess the discrepancy between the confidence level $\rho$ and the corresponding observed frequency $\alpha_{\rho}$. First, a number of confidence levels $\{\rho_j\}_{j=1}^m$ ($0 < \rho_1 < \rho_2 < \dots \rho_m  < 100$) is chosen. For each confidence level $\rho_j$, the corresponding observed frequency $\alpha_{\rho_j}$ can be computed as,
\begin{equation}
    \alpha_{\rho_j} = \vert \{c \vert \text{true}_c \in \hat{\text{CI}}(c, \rho_j), c\in C \} \vert / \vert C \vert \times 100,
\end{equation}
where $\hat{\text{CI}}(c, p_j)$ denotes the generated $\rho_j$\% confidence interval for configuration $c$. Then, the $cal$ score can be computed as,
\begin{equation} \label{eq:cal_score}
    cal = \sum\nolimits_{j=1}^m \omega_j ((\rho_j - \alpha_{\rho_j})/100)^2 \times 100\%,
\end{equation}
with $\{\omega_j\}_{j=1}^m$ being the weights of the metric. In practice, $\omega_j$ is commonly chosen to be 1 for all $j$ \cite{Kuleshov2018}. 

As discussed, a method generates well-calibrated (or reliable) confidence intervals if for all confidence level $\rho, 0\leq \rho \leq 100$, the $\rho\%$ confidence interval contains the ground truths $\rho\%$ of the time, i.e. the confidence level $\rho$ and the corresponding observed frequency $\alpha_{\rho}$ are equal. Thus, a method that generates better confidence intervals will have the $cal$ score lower. A method with the $cal$ score of 0 means that method can generate perfectly calibrated confidence intervals for all confidence levels. In this work, to compute the $cal$ score, we set the weights $\omega_j$ to be 1 (for all $j$) as this is a common choice \cite{Kuleshov2018}. We set the list of confidence levels $\{\rho_j\}_{j=1}^m$ to be 0.05, 0.1, ..., 0.9, 0.95 ($m=19$). This choice of confidence levels is extensive, and thus helps to accurately evaluate the quality of the confidence intervals generated by all methods. Finally, similar to RQ1, we repeat the experiments 20 times, and report the mean and the 95\% confidence interval (margin) of the $cal$ scores. We use t-test with the significant level of 0.05 to statistically compared the $cal$ scores of \textbf{BDLPerf} and \textbf{P4} (with two settings $\Pi_{ho}$ and $\Pi_{he}$).

\paragraph{Results} Table \ref{table:rq2} reports the $cal$ scores of \textbf{BDLPerf} and \textbf{P4} with two settings $\Pi_{ho}$, $\Pi_{he}$ on the 10 subject systems with 3 sampling strategies. With the setting $\Pi_{ho}$, we can see that \textbf{BDLPerf} statistically outperforms \textbf{P4} on 20/30 cases and performs similarly on 6/30 cases. In contrast, with this setting, \textbf{P4} only statistically outperforms \textbf{BDLPerf} on 4/30 cases. The same observation can be concluded for the setting $\Pi_{he}$. That is, \textbf{BDLPerf} statistically outperforms \textbf{P4} on 22/30 cases and performs similarly on 2/30 cases whilst \textbf{P4} only statistically outperforms \textbf{BDLPerf} on 6/30 cases for this setting. Finally, it's worth mentioning that there are various cases when the $cal$ scores of \textbf{P4} are very bad (e.g., the $cal$ scores are higher than 200 or 300\%). On the other hand, the $cal$ scores of \textbf{BDLPerf} are always reasonable across all the subject systems and sampling strategies, specifically, its $cal$ scores are always smaller than 100\%. This is mainly due to our proposed uncertainty calibration technique. It helps to generate well-calibrated confidence intervals across different settings. On the other hand, \textbf{P4} does not have any uncertainty calibration process, and thus, in various cases, it generates poorly-calibrated confidence intervals.

\begin{table*}[htbp]
  \centering
  \caption{
  Experimental results showing the mean and margin values of the $cal$ scores of our proposed method \textbf{BDLPerf} and the basline method \textbf{P4} \cite{dorn2020mastering} with two settings $\Pi_{ho}$, $\Pi_{he}$. The lower the $cal$ scores, the better the methods. Experiments are repeated 20 times. In the column \textit{Better Algorithm}, we report the better method using t-test with p-value of 0.05.
  }
    \begin{tabular}{m{29pt}<{\centering} | m{24pt}<{\centering} | m{19pt}<{\centering}  m{24pt}<{\centering} | m{19pt}<{\centering}  m{24pt}<{\centering} | m{41pt}<{\centering} || m{19pt}<{\centering}  m{24pt}<{\centering} | m{19pt}<{\centering}  m{24pt}<{\centering} | m{41pt}<{\centering} }\hline 
    
      \multirow{2}{*}{System} & \multirow{2}{*}{t-wise} & \multicolumn{2}{c}{BDLPerf} &  \multicolumn{2}{c}{P4($\Pi_{ho}$)} & \multirow{2}{39pt}{\centering Better Algorithm} & \multicolumn{2}{c}{BDLPerf} & \multicolumn{2}{c}{P4($\Pi_{he}$)} & \multirow{2}{39pt}{\centering Better Algorithm}  \\
       &  & Mean & Margin & Mean & Margin & & Mean & Margin & Mean & Margin &  \\ \hline
     \multirow{3}{*}{7z} & 1 & \textbf{104.9} & 21.3 & 123.0 & 1.4 & Same & \textbf{104.9} & 21.3 & 142.4 & 3.2 & BDLPerf \\ 
                         & 2 & \textbf{24.2} & 5.7 & 197.4 & 0.6 & BDLPerf & \textbf{24.2} & 5.7 & 34.1 & 0.5 & BDLPerf \\ 
                         & 3 & \textbf{12.3} & 4.3 & 562.3 & 0.7 & BDLPerf & 12.3 & 4.3 & \textbf{8.2} & 0.1 & Same \\ \hline
  \multirow{3}{*}{BDB-C} & 1 & \textbf{63.0} & 20.3 & 63.9 & 0.5 & Same & \textbf{63.0} & 20.3 & 100.8 & 43.7 & BDLPerf \\ 
                         & 2 & \textbf{27.8} & 8.4 & 53.5 & 1.2 & BDLPerf & \textbf{27.8} & 8.4 & 63.8 & 39.8 & BDLPerf \\ 
                         & 3 & \textbf{8.3} & 2.8 & 420.1 & 3.8 & BDLPerf & \textbf{8.3} & 2.8 & 49.7 & 107.7 & BDLPerf \\ \hline
   \multirow{3}{*}{Dune} & 1 & 11.1 & 3.1 & \textbf{1.6} & 0.1 & P4 & \textbf{11.1} & 3.1 & 27.2 & 1.2 & BDLPerf \\ 
                         & 2 & \textbf{16.8} & 6.9 & 174.6 & 1.3 & BDLPerf & 16.8 & 6.9 & \textbf{0.7} & 0.1 & P4 \\ 
                         & 3 & \textbf{19.6} & 4.7 & 285.8 & 1.9 & BDLPerf & 19.6 & 4.7 & \textbf{2.4} & 0.2 & P4 \\ \hline
 \multirow{3}{*}{Hipacc} & 1 & \textbf{15.7} & 7.3 & 86.3 & 0.7 & BDLPerf & \textbf{15.7} & 7.3 & 141.0 & 1.3 & BDLPerf \\ 
                         & 2 & \textbf{10.0} & 3.1 & 19.8 & 0.2 & BDLPerf & \textbf{10.0} & 3.1 & 148.2 & 0.8 & BDLPerf \\ 
                         & 3 & \textbf{1.5} & 0.4 & 96.2 & 0.4 & BDLPerf & \textbf{1.5} & 0.4 & 18.9 & 0.1 & BDLPerf\\ \hline
   \multirow{3}{*}{LLVM} & 1 & 91.4 & 46.0 & \textbf{2.3} & 0.2 & P4 & 91.4 & 46.0 & \textbf{66.9} & 3.0 & Same \\ 
                         & 2 & \textbf{87.8} & 26.5 & 90.0 & 1.4 & Same & \textbf{87.8} & 26.5 & 181.7 & 1.9 & BDLPerf \\ 
                         & 3 & 39.5 & 17.8 & \textbf{11.7} & 0.5 & P4 & 39.5 & 17.8 & \textbf{4.8} & 0.3 & P4 \\ \hline
  \multirow{3}{*}{Polly} & 1 & \textbf{74.7} & 19.6 & 134.3 & 0.3 & BDLPerf & \textbf{74.7} & 19.6 & 99.9 & 1.9 & BDLPerf \\ 
                         & 2 & 31.7 & 36.4 & \textbf{20.3} & 0.5 & Same & \textbf{31.7} & 36.4 & 119.9 & 1.4 & BDLPerf  \\ 
                         & 3 & \textbf{7.4} & 3.1 & 16.5 & 0.1 & BDLPerf & \textbf{7.4} & 3.1 & 58.6 & 0.2 & BDLPerf \\ \hline
  \multirow{3}{*}{lrzip} & 1 & \textbf{43.3} & 7.8 & 159.8 & 3.5 & BDLPerf & \textbf{43.3} & 7.8 & 75.0 & 4.3 & BDLPerf  \\ 
                         & 2 & \textbf{13.3} & 8.1 & 426.1 & 5.1 & BDLPerf & 13.3 & 8.1 & \textbf{3.4} & 0.9 & P4 \\ 
                         & 3 & \textbf{52.0} & 8.0 & 601.9 & 3.5 & BDLPerf & 52.0 & 8.0 & \textbf{11.4} & 2.7 & P4 \\ \hline
   \multirow{3}{*}{x264} & 1 & 55.5 & 32.5 & \textbf{46.5} & 2.2 & Same & \textbf{55.5} & 32.5 & 405.3 & 67.4 & BDLPerf  \\ 
                         & 2 & 16.7 & 6.1 & \textbf{14.2} & 0.9 & Same & \textbf{16.7} & 6.1 & 49.1 & 2.2 & BDLPerf  \\ 
                         & 3 & \textbf{4.5} & 1.4 & 339.9 & 6.1 & BDLPerf & \textbf{4.5} & 1.4 & 30.7 & 1.2 & BDLPerf  \\ \hline
    \multirow{3}{*}{VP9} & 1 & \textbf{19.6} & 20.5 & 33.1 & 1.2 & BDLPerf & 19.6 & 20.5 & \textbf{3.2} & 9.0 & P4 \\ 
                         & 2 & \textbf{6.1} & 13.6 & 19.0 & 0.6 & BDLPerf & \textbf{6.1} & 13.6 & 14.9 & 24.9 & BDLPerf \\ 
                         & 3 & \textbf{3.3} & 5.8 & 137.6 & 0.7 & BDLPerf & \textbf{3.3} & 5.8 & 28.0 & 55.3 & BDLPerf \\ \hline
 \multirow{3}{*}{JavaGC} & 1 & \textbf{35.9} & 46.3 & 231.9 & 0.6 & BDLPerf & \textbf{35.9} & 46.3 & 325.5 & 1.9 & BDLPerf \\ 
                         & 2 & 33.2 & 28.6 & \textbf{23.1} & 0.3 & P4 & \textbf{33.2} & 28.6 & 202.4 & 1.8 & BDLPerf \\ 
                         & 3 & \textbf{2.4} & 4.2 & 54.5 & 0.3 & BDLPerf & \textbf{2.4} & 4.2 & 46.2 & 4.4 & BDLPerf \\ \hline
    \end{tabular}%
  \label{table:rq2}%
\end{table*}%

% For the subject systems BDB-C, Hipacc, and x264, our method \textbf{BDLPerf} statistically outperforms other baselines for all sampling strategies. For these systems, the $cal$ scores of \textbf{BDLPerf} are much lower than both baselines, i.e., its $cal$ scores are generally two or three times lower. For other subject systems (except Dune), there are few cases when $\Pi_{ho}$ or $\Pi_{he}$ is statistically better, but there are also many cases when \textbf{BDLPerf} is statistically better. Overall, \textbf{BDLPerf} is statistically the best algorithm in 17/24 cases. 

\subsection{RQ3: The Effectiveness of the Proposed Uncertainty Calibration Technique}

In this RQ, we aim to investigate the effectiveness of our proposed uncertainty calibration technique.

\begin{table*}[htbp]
% \fontsize{6}{\baselineskip} \selectfont
  \centering
%   \addtolength{\tabcolsep}{-5pt}
  \caption{Experimental results showing the mean and margin values for the $cal$ scores of our proposed method, \textbf{BDLPerf}, and two methods: \textbf{BDL} and \textbf{BDLPlatt}. The lower the $cal$ scores, the better the methods. Experiments are repeated 20 times. In the column \textit{Best Algorithm}, we report the best method among three methods using a combination of ANOVA and t-test with p-value of 0.05.}
    %\begin{tabular}{ccc|ccc|ccc}
    % \vspace{-2pt}
    \begin{tabular}{m{29pt}<{\centering} | m{24pt}<{\centering} | m{22pt}<{\centering} m{24pt}<{\centering} | m{22pt}<{\centering} m{24pt}<{\centering} | m{22pt}<{\centering} m{24pt}<{\centering} | m{50pt}<{\centering} }\hline 
    
      \multirow{2}{*}{System} & \multirow{2}{*}{t-wise} & \multicolumn{2}{c}{BDLPerf} &  \multicolumn{2}{c}{BDL} & \multicolumn{2}{c}{BDLPlatt} & \multirow{2}{50pt}{\centering Best Algorithm}  \\
       &  & Mean & Margin & Mean & Margin & Mean & Margin & \\ \hline
     \multirow{3}{*}{7z} & 1 & \textbf{104.9} & 21.3 & 247.6 & 65.9 & 285.2 & 65.5 & BDLPerf \\ 
                         & 2 & \textbf{24.2} & 5.7 & 440.0 & 21.6 & 406.5 & 25.4 & BDLPerf \\ 
                         & 3 & \textbf{12.3} & 4.3 & 381.3 & 55.0 & 294.7 & 46.2 & BDLPerf \\ \hline
  \multirow{3}{*}{BDB-C} & 1 & \textbf{63.0} & 20.3 & 149.9 & 35.6 & 219.5 & 48.1 & BDLPerf \\ 
                         & 2 & \textbf{27.8} & 8.4 & 112.5 & 17.8 & 132.9 & 20.7 & BDLPerf \\ 
                         & 3 & \textbf{8.3} & 2.8 & 56.1 & 29.3 & 76.3 & 28.1 & BDLPerf \\ \hline
   \multirow{3}{*}{Dune} & 1 & \textbf{11.1} & 3.1 & 213.8 & 71.1 & 182.7 & 65.8 & BDLPerf \\ 
                         & 2 & \textbf{16.8} & 6.9 & 375.9 & 26.8 & 317.7 & 28.0 & BDLPerf \\ 
                         & 3 & \textbf{19.6} & 4.7 & 196.0 & 45.8 & 136.2 & 40.8 & BDLPerf \\ \hline
 \multirow{3}{*}{Hipacc} & 1 & \textbf{15.7} & 7.3 & 374.7 & 70.1 & 368.3 & 61.7 & BDLPerf \\ 
                         & 2 & \textbf{10.0} & 3.1 & 365.6 & 43.1 & 314.2 & 45.8 & BDLPerf \\ 
                         & 3 & \textbf{1.5} & 0.4 & 173.1 & 49.5 & 140.8 & 83.1 & BDLPerf \\ \hline
   \multirow{3}{*}{LLVM} & 1 & \textbf{91.4} & 46.0 & 252.5 & 61.0 & 326.1 & 56.5 & BDLPerf \\ 
                         & 2 & \textbf{87.8} & 26.5 & 272.3 & 78.5 & 238.0 & 77.6 & BDLPerf \\ 
                         & 3 & \textbf{39.5} & 17.8 & 362.3 & 50.2 & 327.9 & 51.1 & BDLPerf \\ \hline
  \multirow{3}{*}{Polly} & 1 & \textbf{74.7} & 19.6 & 523.9 & 29.8 & 512.0 & 30.0 & BDLPerf \\ 
                         & 2 & \textbf{31.7} & 36.4 & 416.2 & 41.6 & 354.7 & 49.8 & BDLPerf \\ 
                         & 3 & \textbf{7.4} & 3.1 & 433.1 & 34.9 & 245.1 & 187.1 & BDLPerf \\ \hline
  \multirow{3}{*}{lrzip} & 1 & \textbf{43.3} & 7.8 & 133.1 & 36.9 & 163.3 & 35.3 & BDLPerf \\ 
                         & 2 & \textbf{13.3} & 8.1 & 36.6 & 16.0 & 27.7 & 9.6 & Same \\ 
                         & 3 & 52.0 & 8.0 & 30.8 & 23.1 & \textbf{26.5} & 13.3 & Same \\ \hline
   \multirow{3}{*}{x264} & 1 & \textbf{55.5} & 32.5 & 207.7 & 55.3 & 243.3 & 52.9 & BDLPerf \\ 
                         & 2 & \textbf{16.7} & 6.1 & 254.4 & 51.0 & 236.9 & 51.9 & BDLPerf \\ 
                         & 3 & \textbf{4.5} & 1.4 & 130.0 & 44.8 & 81.3 & 38.0 & BDLPerf \\ \hline
    \multirow{3}{*}{VP9} & 1 & \textbf{19.6} & 20.5 & 55.5 & 204.5 & 70.4 & 206.1 & BDLPerf \\ 
                         & 2 & \textbf{6.1} & 13.6 & 367.6 & 123.0 & 369.7 & 122.8 & BDLPerf \\ 
                         & 3 & \textbf{3.3} & 5.8 & 315.3 & 263.9 & 314.6 & 266.5 & BDLPerf \\ \hline
 \multirow{3}{*}{JavaGC} & 1 & \textbf{35.9} & 46.3 & 385.3 & 383.7 & 405.6 & 368.0 & BDLPerf \\
                         & 2 & \textbf{33.2} & 28.6 & 489.6 & 135.8 & 490.6 & 141.5 & BDLPerf \\
                         & 3 & \textbf{2.4} & 4.2 & 359.4 & 269.9 & 354.8 & 278.6 & BDLPerf \\ \hline
    \end{tabular}%
  \label{table:rq3}%
\end{table*}%

\paragraph{Setup} We use the same setup as in RQ2. We compare \textbf{BDLPerf} with two methods: one directly uses the generated confidence intervals by the BNNs (namely \textbf{BDL}), and one uses the generated confidence intervals by the BNNs combined with the Platt scaling calibration technique \cite{Kuleshov2018} (namely \textbf{BDLPlatt}). We compare the three methods using the $cal$ score (Eq. \ref{eq:cal_score}) to evaluate their effectiveness in generating confidence intervals for configurable software performance. As we have three different methods, we first perform an ANOVA test with p-value of 0.05 to see if there is any method that is significantly different compared to the other two methods. After that, we perform t-test pairwise (with p-value of 0.05) and choose the best method as the method that is statistically significantly better than both other two methods.

\paragraph{Results} In Table \ref{table:rq3}, we report the $cal$ scores of the three methods. \textbf{BDLPerf} statistically outperforms the other two methods on all subject systems and sampling strategies (except lrzip). Besides, its $cal$ scores are lower than the $cal$ scores of other two methods by a very high margin (the lower the $cal$ scores, the better the method). The results of \textbf{BDL} demonstrate that without uncertainty calibration, the generated confidence intervals can be very unreliable. On the other hand, even with standard uncertainty calibration (\textbf{BDLPlatt}), the generated confidence intervals are still largely unreliable. This is mostly due to the issue of limited training data of the configurable software performance prediction problem. Our proposed technique \textbf{BDLPerf} overcomes the limitations of the standard uncertainty calibration technique, and performs well for all the subject systems and sampling strategies.

\begin{table}[]
  \centering
  \caption{Average time cost (in minutes) of \textbf{BDLPerf} over 20 experiments for each combination of subject system and sampling strategy. All the time cost is measured when running \textbf{BDLPerf} on a Windows 10 computer with Intel Core i5-8265U CPU 1.6GHz 8GB RAM (no GPU).}
%   \vspace{-2pt}
    \begin{tabular}{m{23pt}<{\centering} | m{29pt}<{\centering} | m{52pt}<{\centering} | m{29pt}<{\centering} | m{52pt}<{\centering}  }\hline
    t-wise & System & Time cost & System & Time cost \\ \hline
    %   \multirow{2}{*}{System} & \multirow{2}{*}{t-wise} & \multirow{2}{*}{Time cost}   \\ \hline
     1 & \multirow{3}{*}{7z} & 36.32 $\pm$ 4.62 & \multirow{3}{*}{LLVM} & 36.88 $\pm$ 4.21 \\ 
     2 &                     & 109.15 $\pm$ 8.01 & & 26.79 $\pm$ 2.34 \\ 
     3 &                     & 738.03 $\pm$ 97.49 & & 38.73 $\pm$ 5.98 \\ \hline
     1 & \multirow{3}{*}{BDB-C} & 21.31 $\pm$ 2.80 & \multirow{3}{*}{Polly} & 32.34 $\pm$ 1.66 \\ 
     2 &                        & 25.15 $\pm$ 2.03 & & 97.24 $\pm$ 10.57 \\ 
     3 &                        & 45.01 $\pm$ 4.37 & & 574.17 $\pm$ 260.1 \\ \hline
     1 & \multirow{3}{*}{Dune} & 37.94 $\pm$ 6.60 & \multirow{3}{*}{lrzip} & 21.06 $\pm$ 3.93 \\ 
     2 &                       & 48.78 $\pm$ 5.94 & & 27.61 $\pm$ 2.39 \\ 
     3 &                       & 145.68 $\pm$ 21.86 & & 36.90 $\pm$ 1.89 \\ \hline
     1 & \multirow{3}{*}{Hipacc} & 47.47 $\pm$ 2.97 & \multirow{3}{*}{x264} & 21.35 $\pm$ 4.17\\ 
     2 &                         & 149.07 $\pm$ 42.15 & & 37.75 $\pm$ 2.81 \\ 
     3 &                         & 792.44 $\pm$ 109.5 & & 36.09 $\pm$ 2.79 \\ \hline
     1 & \multirow{3}{*}{VP9 } & 33.02 $\pm$ 2.88 & \multirow{3}{*}{JavaGC} & 36.53 $\pm$ 4.11 \\ 
     2 &                         & 86.43 $\pm$ 8.38 & & 66.46 $\pm$ 5.73 \\ 
     3 &                         & 706.60 $\pm$ 207.6 & & 516.35 $\pm$ 53.6 \\ \hline
    \end{tabular}%
  \label{table:rq4}%
%   \vspace{-6pt}
\end{table}%

\subsection{RQ4: Time Cost of \textbf{BDLPerf}}

In this RQ, we discuss the time cost of \textbf{BDLPerf} to evaluate its efficiency. In Table \ref{table:rq4}, we report the average running time (over 20 experiments) of \textbf{BDLPerf} for each combination of subject system and sampling strategy. The running time includes the time to tune the hyperparameters, train the BNNs, generate the scalar predictions and confidence intervals, and calibrate the confidence intervals. All the time cost is measured when running \textbf{BDLPerf} on a Windows 10 computer with Intel Core i5-8265U CPU 1.6GHz 8GB RAM (no GPU).
% \hy{need to report the average runtime cost of x runs?}

It can be seen that the running time of \textbf{BDLPerf} is reasonable. For subject systems with the number of configuration options $n$ less than 20 (e.g., BDB-C, LLVM, lrzip, x264), the average running time of \textbf{BDLPerf} for each subject system and sampling strategy is from 20 minutes to 40 minutes. For subject systems with the number of configuration options more than 20 (e.g., 7z, Dune, Hipacc, Polly, VP9, JavaGC), the average running time of \textbf{BDLPerf} is from 30 minutes to 12 hours. These average running time is reasonable, demonstrating the efficiency of our hyperparameter tuning strategy.

It is worth noting that the running time of \textbf{P4} are generally faster. For subject systems with less than 20 configuration options, using the same computer, it takes around 1-3 minutes to train the model and generate the confidence interval. For large subject systems with more than 20 configuration options, it takes approximately 2 minutes to 4 hours to train the model and generate the scalar predictions and confidence intervals.

% Our proposed approach, $BDLPerf$, runs slower compared to $\pi_{ho}$ and $\pi_{he}$. In particular, for small configurable software systems (i.e., systems with less than 15 configuration options), it takes approximately 20-30 minutes to tune the hyperparameters, train the model, and generate the uncertainty. For large configurable software systems, it takes 30 minutes to 2 hours to tune the hyperparameters, train the model, and generate the uncertainty. Note that the time cost of our proposed approach is higher compared to existing approaches, however, it is reasonable, given the win on the performance. Especially, for large configurable software systems, the time cost to construct the training datasets is much more expensive compared to the time cost of tuning the hyperparameters and train the Bayesian deep learning model.

%\section{Discussion}

%\hy{can also discuss why the proposed method works}

%\hy{can also add a Discussion section to discuss certain things like Why it Works, Possible Application Scenarios, etc}
\section{Threats to Validity}

Similar to existing work in predicting performance of configurable software systems, an internal threat to validity can arise from the measurement error bias. In this work, we use the benchmark datasets from previous work which performed careful measurement, and also control this bias by repeating the measurements multiple times \cite{Kaltenecker2020samplingdistance, dorn2020mastering}. Another internal threat of validity might come from the stochastic behavior of the package TensorFlow Probability 0.15.0 \cite{TFProbability} which we use to implement our proposed method. To avoid this threat, for each combination of subject system and t-wise sampling strategy, we repeat the experiments 20 times, and report both the average and the 95\% confidence interval (margin) of the results. We also use hypothesis testing methods (e.g., t-test and ANOVA) to evaluate our experimental results. Finally, an internal threat of validity can arise from the fact that the measurements of the subject systems we use do not take into account the varying workload, i.e., the workload of the systems are assumed to be fixed for all configurations \cite{Kaltenecker2020samplingdistance, dorn2020mastering}. In practice, different workload may result into different performance measurements for the same configuration \cite{Lesoil2021PerfInput}. However, these workload can be feasibly incorporated into the modeling process, so we assume this threat is minimal.

% on estimating the uncertainty of software performance prediction, the internal threat to validity can arise from measurement error bias. To avoid this type of issue, for all subject systems, we repeat the experiments 20 times. Another thread of validity might come from the package tensorflow probability. We use this package to construct our Bayesian deep learning model. Based on our experience, the package is well-known in the community, and is validated and tested extensively by the development team\hy{then why it is a threat}.

For the external threat of validity, we evaluate all the methods including our proposed method on a wide range of configurable software systems. These subject systems range from small-scale to large-scale systems, with the configuration options including both binary and numeric options. They have different characteristics and are from different application domains. These systems have been used extensively in the literature to evaluate performance prediction for configurable software systems \cite{siegmund2015performance, Zhang2015PerfFL, Guo2018Decart, ha2019deepperf, ha2019performance, dorn2020mastering}. Even though these systems may not be representative to all types of configurable software systems in real world, we believe that our proposed method can be applicable for a wide range of configurable software systems.

\section{Related Work}

The problem of predicting the performance of configurable software systems has attracted attention of software engineering researchers. There are numerous research work aiming to develop novel methods for estimating performance of configurable software systems from a limited number of measurement data. Siegmund et al. \cite{siegmund2015performance} proposes to use linear regression combined with different sampling techniques to build a performance influence model for configurable software systems. Guo et al. \cite{GuoCART2013} suggests to use decision trees to model configurable software performance, and later it was extended to combine with various machine learning hyperparameter tuning technique to become more data-efficient \cite{Guo2018Decart}. Zhang et al. \cite{Zhang2015PerfFL} uses Fourier learning as the learning methodology whilst also providing a theoretical analysis on the accuracy of the proposed model. Recently, Ha et al. \cite{ha2019deepperf} proposes to use deep sparse neural network to model the configurable software system performance. Whilst these work can give good scalar predictions for the configurable software systems performance, their limitation is that they cannot provide the confidence intervals of these scalar predictions.

In terms of generating the confidence intervals for configurable software performance, Dorn et al. \cite{dorn2020mastering} tackled this problem for the first time. The key idea is to use a Lasso regression model to model the relationship between the configuration options and the system performance. Bayesian inference is then performed via probabilistic programming to infer the posterior distribution of the model's parameters, thus, infer the posterior distribution of the scalar prediction. Our work also aims to generate confidence intervals of configurable software performance, however, we use deep neural networks to model the configurable software performance. We also suggest a new technique for uncertainty calibration of the generated confidence intervals. We have compared the two methods in Section \ref{sec:exp-results}. Our experimental results show that our method not only can generate better scalar predictions, but also the confidence intervals of these scalar predictions.

%  We have compared it with our proposed  Bayesian Deep Learning based method (Section X). % to generate uncertainty. %Apart from these work, there are some work aiming to solve other issues of configurable software systems.

The work in \cite{Jamshidi2017transferlearningperf} investigates a new method of transferring knowledge among similar configurable software systems, thus helps to improve the accuracy of the prediction model whilst only requiring a minimal number of data. The work in \cite{Nair2020} proposes a sequential model-based method to find the configuration with the highest/lowest performance value using the least number of measurement data. In \cite{Sarkar2015Sampling}, a new sampling method is proposed to select the measurement data in a cost-efficient manner whilst still achieving high model prediction accuracy. The work in \cite{Kaltenecker2020samplingdistance} devises a distance-based sampling strategy based on a distance metric and the configurations' probability distribution across the configuration. More recently, there are new approaches aiming to construct the prediction models of configurable software systems in a white-box manner, that is, using some internal information of the software systems to aid with the performance prediction modeling process \cite{Weber2021Whitebox,Velez2021Whitebox}. Unlike the above work, we aim to build a prediction model that can provide both scalar predictions and their confidence intervals in a black-box manner.

% Another work in (cite) investigation the performance of configurable software systems among different versions of the software systems. 

% \vspace{-12pt}
\section{Conclusion}

In this paper, we have proposed a Bayesian deep learning based method, namely \textbf{BDLPerf}, for providing both scalar predictions and confidence intervals for the performance of configurable software systems. In particular, we propose to use deep neural networks to model the configurable software performance, and use the Variational Inference method combined with the Ensemble method to generate the confidence intervals of the software performance. We further develop a novel technique based on the Platt scaling technique to calibrate the confidence intervals that work well in the limited training data scenario. Finally, we propose a strategy to efficiently identify the optimal hyperparameters of our proposed model within a reasonable amount of time. Our experimental results on ten real-world configurable software systems confirm the effectiveness and efficiency of our proposed method.

Our code and data are available at: \url{https://github.com/HuongHa12/BDLPerf}. %A GitHub repository (Python) will be made publicly available after the paper is accepted.

% \hy{Our code and experimental results are publicly available at: http:// Github page}

% \begin{acks}
% % To Robert, for the bagels and explaining CMYK and color spaces.
% \end{acks}

%\bibliographystyle{ACM-Reference-Format}
\bibliographystyle{IEEEtran}
\bibliography{reference}

\begin{thebibliography}{10}
\providecommand{\url}[1]{#1}
\csname url@samestyle\endcsname
\providecommand{\newblock}{\relax}
\providecommand{\bibinfo}[2]{#2}
\providecommand{\BIBentrySTDinterwordspacing}{\spaceskip=0pt\relax}
\providecommand{\BIBentryALTinterwordstretchfactor}{4}
\providecommand{\BIBentryALTinterwordspacing}{\spaceskip=\fontdimen2\font plus
\BIBentryALTinterwordstretchfactor\fontdimen3\font minus
  \fontdimen4\font\relax}
\providecommand{\BIBforeignlanguage}[2]{{%
\expandafter\ifx\csname l@#1\endcsname\relax
\typeout{** WARNING: IEEEtran.bst: No hyphenation pattern has been}%
\typeout{** loaded for the language `#1'. Using the pattern for}%
\typeout{** the default language instead.}%
\else
\language=\csname l@#1\endcsname
\fi
#2}}
\providecommand{\BIBdecl}{\relax}
\BIBdecl

\bibitem{Han2016PerfEmpirical}
\BIBentryALTinterwordspacing
X.~Han and T.~Yu, ``An empirical study on performance bugs for highly
  configurable software systems,'' in \emph{Proceedings of the 10th {ACM/IEEE}
  International Symposium on Empirical Software Engineering and Measurement,
  {ESEM} 2016, Ciudad Real, Spain, September 8-9, 2016}.\hskip 1em plus 0.5em
  minus 0.4em\relax {ACM}, 2016, pp. 23:1--23:10. [Online]. Available:
  \url{https://doi.org/10.1145/2961111.2962602}
\BIBentrySTDinterwordspacing

\bibitem{Velez2022PerfDebug}
M.~Velez, P.~Jamshidi, N.~Siegmund, S.~Apel, and C.~K{\"{a}}stner, ``On
  debugging the performance of configurable software systems: Developer needs
  and tailored tool support,'' in \emph{44th {IEEE/ACM} 44th International
  Conference on Software Engineering, {ICSE} 2022, Pittsburgh, PA, USA, May
  25-27, 2022}.\hskip 1em plus 0.5em minus 0.4em\relax {ACM}, 2022, pp.
  1571--1583.

\bibitem{He2020PerfBugs}
H.~He, Z.~Jia, S.~Li, E.~Xu, T.~Yu, Y.~Yu, J.~Wang, and X.~Liao, ``Cp-detector:
  Using configuration-related performance properties to expose performance
  bugs,'' in \emph{35th {IEEE/ACM} International Conference on Automated
  Software Engineering, {ASE} 2020, Melbourne, Australia, September 21-25,
  2020}.\hskip 1em plus 0.5em minus 0.4em\relax {IEEE}, 2020, pp. 623--634.

\bibitem{Li2016EngeryOptimize}
D.~Li, Y.~Lyu, J.~Gui, and W.~G.~J. Halfond, ``Automated energy optimization of
  {HTTP} requests for mobile applications,'' in \emph{Proceedings of the 38th
  International Conference on Software Engineering, {ICSE} 2016, Austin, TX,
  USA, May 14-22, 2016}, L.~K. Dillon, W.~Visser, and L.~A. Williams,
  Eds.\hskip 1em plus 0.5em minus 0.4em\relax {ACM}, 2016, pp. 249--260.

\bibitem{Song2017PerfDiagnosis}
L.~Song and S.~Lu, ``Performance diagnosis for inefficient loops,'' in
  \emph{Proceedings of the 39th International Conference on Software
  Engineering, {ICSE} 2017, Buenos Aires, Argentina, May 20-28, 2017},
  S.~Uchitel, A.~Orso, and M.~P. Robillard, Eds.\hskip 1em plus 0.5em minus
  0.4em\relax {IEEE} / {ACM}, 2017, pp. 370--380.

\bibitem{Wilke2013EngeryMobileApps}
C.~Wilke, S.~Richly, S.~G{\"{o}}tz, C.~Piechnick, and U.~A{\ss}mann, ``Energy
  consumption and efficiency in mobile applications: {A} user feedback study,''
  in \emph{2013 {IEEE} International Conference on Green Computing and
  Communications (GreenCom) and {IEEE} Internet of Things (iThings) and {IEEE}
  Cyber, Physical and Social Computing (CPSCom), Beijing, China, August 20-23,
  2013}.\hskip 1em plus 0.5em minus 0.4em\relax {IEEE}, 2013, pp. 134--141.

\bibitem{dorn2020mastering}
J.~Dorn, S.~Apel, and N.~Siegmund, ``Mastering uncertainty in performance
  estimations of configurable software systems,'' in \emph{Proceedings of the
  35th IEEE/ACM International Conference on Automated Software Engineering
  (ASE)}, Melbourne, 2020, pp. 684--696.

\bibitem{siegmund2015performance}
N.~Siegmund, A.~Grebhahn, S.~Apel, and C.~K{\"a}stner, ``Performance-influence
  models for highly configurable systems,'' in \emph{Proceedings of the 2015
  10th Joint Meeting on Foundations of Software Engineering (FSE)}, 2015, pp.
  284--294.

\bibitem{Zhang2015PerfFL}
Y.~Zhang, J.~Guo, E.~Blais, and K.~Czarnecki, ``Performance prediction of
  configurable software systems by fourier learning {(T)},'' in
  \emph{Proceedings of the 30th {IEEE/ACM} International Conference on
  Automated Software Engineering (ASE)}, 2015, pp. 365--373.

\bibitem{Guo2018Decart}
J.~Guo, D.~Yang, N.~Siegmund, S.~Apel, A.~Sarkar, P.~Valov, K.~Czarnecki,
  A.~Wasowski, and H.~Yu, ``Data-efficient performance learning for
  configurable systems,'' \emph{Empirical Software Engineering}, vol.~23,
  no.~3, pp. 1826--1867, 2018.

\bibitem{ha2019deepperf}
H.~Ha and H.~Zhang, ``Deepperf: performance prediction for configurable
  software with deep sparse neural network,'' in \emph{Proceedings of the 41st
  IEEE/ACM International Conference on Software Engineering (ICSE)}, 2019, pp.
  1095--1106.

\bibitem{ha2019performance}
------, ``Performance-influence model for highly configurable software with
  fourier learning and lasso regression,'' in \emph{Proceedings of the 2019
  IEEE International Conference on Software Maintenance and Evolution (ICSME)},
  2019, pp. 470--480.

\bibitem{KennerPerfSafety2021}
A.~Kenner, R.~May, J.~Kr{\"{u}}ger, G.~Saake, and T.~Leich, ``Safety, security,
  and configurable software systems: a systematic mapping study,'' in
  \emph{{SPLC} '21: 25th {ACM} International Systems and Software Product Line
  Conference, Leicester, United Kingdom, September 6-11, 2021, Volume {A}},
  M.~R. Mousavi and P.~Schobbens, Eds.\hskip 1em plus 0.5em minus 0.4em\relax
  {ACM}, 2021, pp. 148--159.

\bibitem{Gal2016Uncertainty}
Y.~Gal, ``Uncertainty in deep learning,'' Ph.D. dissertation, University of
  Cambridge, 2016.

\bibitem{Gal2016MCdropout}
Y.~Gal and Z.~Ghahramani, ``Dropout as a bayesian approximation: Representing
  model uncertainty in deep learning,'' in \emph{Proceedings of the 33rd
  International Conference on International Conference on Machine Learning
  (ICML)}, 2016, p. 1050–1059.

\bibitem{AbdarUncertainty2021}
M.~Abdar, F.~Pourpanah, S.~Hussain, D.~Rezazadegan, L.~Liu, M.~Ghavamzadeh,
  P.~W. Fieguth, X.~Cao, A.~Khosravi, U.~R. Acharya, V.~Makarenkov, and
  S.~Nahavandi, ``A review of uncertainty quantification in deep learning:
  Techniques, applications and challenges,'' \emph{Inf. Fusion}, vol.~76, pp.
  243--297, 2021.

\bibitem{Lakshminarayanan2017}
B.~Lakshminarayanan, A.~Pritzel, and C.~Blundell, ``Simple and scalable
  predictive uncertainty estimation using deep ensembles,'' in \emph{Advances
  in Neural Information Processing Systems 30 (NeurIPS)}, 2017, pp. 6402--6413.

\bibitem{HullermeierUncertainty2021}
\BIBentryALTinterwordspacing
E.~H{\"{u}}llermeier and W.~Waegeman, ``Aleatoric and epistemic uncertainty in
  machine learning: an introduction to concepts and methods,'' \emph{Mach.
  Learn.}, vol. 110, no.~3, pp. 457--506, 2021. [Online]. Available:
  \url{https://doi.org/10.1007/s10994-021-05946-3}
\BIBentrySTDinterwordspacing

\bibitem{Goodfellow2016DL}
I.~Goodfellow, Y.~Bengio, and A.~Courville, \emph{Deep Learning}.\hskip 1em
  plus 0.5em minus 0.4em\relax MIT Press, 2016,
  \url{http://www.deeplearningbook.org}.

\bibitem{Barron1993}
A.~R. Barron, ``Universal approximation bounds for superpositions of a
  sigmoidal function,'' \emph{IEEE Transactions on Information Theory},
  vol.~39, no.~3, pp. 930--945, 1993.

\bibitem{Hornik1989}
K.~Hornik, M.~Stinchcombe, and H.~White, ``Multilayer feedforward networks are
  universal approximators,'' \emph{Neural Networks}, vol.~2, no.~5, pp.
  359--366, 1989.

\bibitem{Hornik1991}
K.~Hornik, ``Approximation capabilities of multilayer feedforward networks,''
  \emph{Neural Networks}, vol.~4, no.~2, pp. 251--257, 1991.

\bibitem{Funahashi1989}
K.-I. Funahashi, ``On the approximate realization of continuous mappings by
  neural networks,'' \emph{Neural Networks}, vol.~2, no.~3, pp. 183--192, 1989.

\bibitem{Jordan1999VI}
M.~I. Jordan, Z.~Ghahramani, T.~S. Jaakkola, and L.~K. Saul, ``An introduction
  to variational methods for graphical models,'' \emph{Machine Learning},
  vol.~37, no.~2, pp. 183--233, 1999.

\bibitem{Mackay1992}
D.~MacKay, ``A practical bayesian framework for backpropagation networks,''
  \emph{Neural Computation}, vol.~4, no.~3, p. 448–472, May 1992.

\bibitem{Neal1995}
R.~Neal, ``Bayesian learning for neural networks,'' Ph.D. dissertation,
  University of Toronto, 1995.

\bibitem{Jamshidi2017transferlearningperf}
P.~Jamshidi, N.~Siegmund, M.~Velez, C.~K{\"{a}}stner, A.~Patel, and Y.~Agarwal,
  ``Transfer learning for performance modeling of configurable systems: an
  exploratory analysis,'' in \emph{Proceedings of the 32nd {IEEE/ACM}
  International Conference on Automated Software Engineering (ASE)}, 2017, pp.
  497--508.

\bibitem{Siegmund2012performanceauto}
N.~Siegmund, S.~S. Kolesnikov, C.~K{\"{a}}stner, S.~Apel, D.~S. Batory,
  M.~Rosenm{\"{u}}ller, and G.~Saake, ``Predicting performance via automated
  feature-interaction detection,'' in \emph{34th International Conference on
  Software Engineering (ICSE)}, 2012, pp. 167--177.

\bibitem{Kuleshov2018}
V.~Kuleshov, N.~Fenner, and S.~Ermon, ``Accurate uncertainties for deep
  learning using calibrated regression,'' in \emph{Proceedings of the 35th
  International Conference on Machine Learning (ICML)}, vol.~80, 2018, pp.
  2801--2809.

\bibitem{Platt99scaling}
J.~C. Platt, ``Probabilistic outputs for support vector machines and
  comparisons to regularized likelihood methods,'' in \emph{Advances in Large
  Margin Classifiers}.\hskip 1em plus 0.5em minus 0.4em\relax MIT Press, 1999,
  pp. 61--74.

\bibitem{Snoek2012}
J.~Snoek, H.~Larochelle, and R.~Adams, ``Practical bayesian optimization of
  machine learning algorithms,'' in \emph{Proceedings of the 25th International
  Conference on Neural Information Processing Systems (NIPS)}, USA, 2012, pp.
  2951--2959.

\bibitem{gurevich2019pairing}
P.~Gurevich and H.~Stuke, ``Pairing an arbitrary regressor with an artificial
  neural network estimating aleatoric uncertainty,'' \emph{Neurocomputing},
  vol. 350, pp. 291--306, 2019.

\bibitem{hullermeier2021aleatoric}
E.~H{\"u}llermeier and W.~Waegeman, ``Aleatoric and epistemic uncertainty in
  machine learning: An introduction to concepts and methods,'' \emph{Machine
  Learning}, vol. 110, no.~3, pp. 457--506, 2021.

\bibitem{Kendall2017uncertainty}
A.~Kendall and Y.~Gal, ``What uncertainties do we need in bayesian deep
  learning for computer vision?'' in \emph{Advances in Neural Information
  Processing Systems 30 (NeurIPS)}, 2017, pp. 5574--5584.

\bibitem{Duane1987MCMC}
S.~Duane, A.~Kennedy, B.~J. Pendleton, and D.~Roweth, ``Hybrid monte carlo,''
  \emph{Physics Letters B}, vol. 195, no.~2, pp. 216--222, 1987.

\bibitem{Lakshminarayanan17}
B.~Lakshminarayanan, A.~Pritzel, and C.~Blundell, ``Simple and scalable
  predictive uncertainty estimation using deep ensembles,'' in \emph{Advances
  in Neural Information Processing Systems 30 (NeurIPS)}, 2017, pp. 6402--6413.

\bibitem{Abdar2021UncertaintyRv}
M.~Abdar, F.~Pourpanah, S.~Hussain, D.~Rezazadegan, L.~Liu, M.~Ghavamzadeh,
  P.~Fieguth, X.~Cao, A.~Khosravi, U.~R. Acharya, V.~Makarenkov, and
  S.~Nahavandi, ``A review of uncertainty quantification in deep learning:
  Techniques, applications and challenges,'' \emph{Information Fusion},
  vol.~76, pp. 243--297, 2021.

\bibitem{Tibshirani1996lasso}
R.~Tibshirani, ``Regression shrinkage and selection via the lasso,''
  \emph{Journal of the Royal Statistical Society. Series B (Methodological)},
  vol.~58, no.~1, pp. 267--288, 1996.

\bibitem{Guo2017calibration}
C.~Guo, G.~Pleiss, Y.~Sun, and K.~Q. Weinberger, ``On calibration of modern
  neural networks,'' in \emph{Proceedings of the 34th International Conference
  on Machine Learning (ICML)}, 2017, p. 1321–1330.

\bibitem{Jones1998}
D.~Jones, M.~Schonlau, and W.~Welch, ``Efficient global optimization of
  expensive black-box functions,'' \emph{Journal of Global Optimization},
  vol.~13, no.~4, pp. 455--492, Dec. 1998.

\bibitem{Mockus1978}
J.~Mockus, V.~Tiesis, and A.~Zilinskas, ``The application of bayesian methods
  for seeking the extremum,'' \emph{Toward Global Optimization}, vol.~2, no.
  117-129, p.~2, 1978.

\bibitem{Shahriari16a}
B.~Shahriari, K.~Swersky, Z.~Wang, R.~P. Adams, and N.~de~Freitas, ``Taking the
  human out of the loop: A review of bayesian optimization,'' \emph{Proceedings
  of the IEEE}, vol. 104, no.~1, pp. 148--175, 2016.

\bibitem{Turner2021BO}
R.~Turner, D.~Eriksson, M.~McCourt, J.~Kiili, E.~Laaksonen, Z.~Xu, and
  I.~Guyon, ``Bayesian optimization is superior to random search for machine
  learning hyperparameter tuning: Analysis of the black-box optimization
  challenge 2020,'' in \emph{Proceedings of the NeurIPS 2020 Competition and
  Demonstration Track}, vol. 133, 06--12 Dec 2021, pp. 3--26.

\bibitem{TFProbability}
``{TensorFlow Probability},'' \url{https://www.tensorflow.org/probability},
  accessed 2022-05-06.

\bibitem{Kingma2015Adam}
D.~P. Kingma and J.~Ba, ``Adam: {A} method for stochastic optimization,'' in
  \emph{Proceedings of the 3rd International Conference on Learning
  Representations (ICLR), San Diego, CA, USA}, 2015.

\bibitem{Kullback1951KLdivergence}
S.~Kullback and R.~A. Leibler, ``On information and sufficiency,'' \emph{The
  Annals of Mathematical Statistics}, vol.~22, no.~1, pp. 79--86, 1951.

\bibitem{gpyopt2016}
T.~G. authors, ``{GPyOpt}: A bayesian optimization framework in python,''
  \url{http://github.com/SheffieldML/GPyOpt}, 2016.

\bibitem{Kaltenecker2020samplingdistance}
C.~Kaltenecker, A.~Grebhahn, N.~Siegmund, J.~Guo, and S.~Apel, ``Distance-based
  sampling of software configuration spaces,'' in \emph{Proceedings of the 41st
  IEEE/ACM International Conference on Software Engineering (ICSE)}, 2019, pp.
  1084--1094.

\bibitem{P4Uncertainty}
``{P4 project page},''
  \url{https://github.com/AI-4-SE/Mastering-Uncertainty-in-Performance-Estimations-of-Configurable-Software-Systems},
  accessed 2022-05-06.

\bibitem{Lesoil2021PerfInput}
\BIBentryALTinterwordspacing
L.~Lesoil, M.~Acher, A.~Blouin, and J.-M. Jézéquel, ``The interaction between
  inputs and configurations fed to software systems: an empirical study,''
  2021. [Online]. Available: \url{https://arxiv.org/abs/2112.07279}
\BIBentrySTDinterwordspacing

\bibitem{GuoCART2013}
J.~Guo, K.~Czarnecki, S.~Apel, N.~Siegmund, and A.~Wasowski,
  ``Variability-aware performance prediction: {A} statistical learning
  approach,'' in \emph{2013 28th {IEEE/ACM} International Conference on
  Automated Software Engineering, {ASE} 2013, Silicon Valley, CA, USA, November
  11-15, 2013}, E.~Denney, T.~Bultan, and A.~Zeller, Eds.\hskip 1em plus 0.5em
  minus 0.4em\relax {IEEE}, 2013, pp. 301--311.

\bibitem{Nair2020}
V.~Nair, Z.~Yu, T.~Menzies, N.~Siegmund, and S.~Apel, ``Finding faster
  configurations using {FLASH},'' \emph{IEEE Transactions on Software
  Engineering}, vol.~46, no.~7, pp. 794--811, 2020.

\bibitem{Sarkar2015Sampling}
A.~Sarkar, J.~Guo, N.~Siegmund, S.~Apel, and K.~Czarnecki, ``Cost-efficient
  sampling for performance prediction of configurable systems {(T)},'' in
  \emph{Proceedings of the 30th {IEEE/ACM} International Conference on
  Automated Software Engineering (ASE)}, 2015, pp. 342--352.

\bibitem{Weber2021Whitebox}
M.~Weber, S.~Apel, and N.~Siegmund, ``White-box performance-influence models:
  {A} profiling and learning approach,'' in \emph{Proceedings of the 43rd
  {IEEE/ACM} International Conference on Software Engineering (ICSE)}, 2021,
  pp. 1059--1071.

\bibitem{Velez2021Whitebox}
M.~Velez, P.~Jamshidi, N.~Siegmund, S.~Apel, and C.~K{\"{a}}stner, ``White-box
  analysis over machine learning: Modeling performance of configurable
  systems,'' in \emph{Proceedings of the 43rd {IEEE/ACM} International
  Conference on Software Engineering (ICSE)}, 2021, pp. 1072--1084.

\end{thebibliography}

% \appendix









\end{document}